%% file: batiscafo.tex
\documentclass[10pt,a4paper]{article}
\usepackage{mathrsfs} 
\usepackage[T1]{fontenc}
\usepackage{comment}
\usepackage{placeins}
\usepackage[utf8]{inputenc}
\usepackage{amssymb,amsmath,amsthm,amsfonts}    
\usepackage{bm}                                 
\usepackage{mathtools}                          
\usepackage{stmaryrd}                           
\usepackage{breqn}                              
\usepackage[textfont=footnotesize,labelfont={footnotesize,bf}]{caption}
\usepackage{subcaption}
\usepackage{booktabs}
\usepackage[inline]{enumitem}
\usepackage{graphicx}
\usepackage[dvipsnames]{xcolor}
\usepackage[top=30mm,right=30mm,left=30mm,bottom=30mm]{geometry}
\usepackage[colorlinks=true,linkcolor=blue,urlcolor=blue,citecolor=green]{hyperref}
\usepackage[affil-it,auth-sc]{authblk}
\usepackage[colorinlistoftodos,textwidth=25mm,textsize=footnotesize]{todonotes}
\usepackage{lipsum}                             
\usepackage[
	backend=biber,
	style=numeric,
	citestyle=numeric-comp,
        uniquename=init,
	maxbibnames=99,
	minbibnames=99,
	maxcitenames=2,
	mincitenames=1,
	giveninits=true,
	sorting=none,
	sortlocale=auto,
	natbib=true,
	url=false,
	isbn=false,
	doi=true,
	eprint=true
]{biblatex}
\input{definitions}                             
\graphicspath{{./}{./figures/}}                 

\title{The circular disc made of linear elastic incompressible material and the \lq bathyscaphe lesson'
}

\author[1]{D. Bigoni\footnote{Corresponding author: e-mail: \href{mailto:bigoni@ing.unitn.it}{bigoni@ing.unitn.it}; phone: +39\,0461\,282507.}}
\author[2]{S.G. Mogilevskaya}
\author[1]{A. Piccolroaz}
\author[3]{M. Gaibotti}

\affil[1]{Instabilities Lab, University of Trento, Trento, Italy}
\affil[2]{Department of Civil, Environmental and Geo-Engineering, University of Minnesota, 500 Pillsbury Drive S.E., Minneapolis, MN 55455-0116, USA}
\affil[3]{SISSA--International School for Advanced Studies, via Bonomea 265, 34136 Trieste, Italy}

\begin{document}

\date{Dedicated to the memory of Professor A.P.S. Selvadurai} 

\maketitle

\begin{abstract}
\noindent 
A linear elastic circular disc is analyzed under a self-equilibrated system of loads applied along its boundary. A distinctive feature of the investigation, conducted using complex variable analysis, is the assumption that the material is incompressible (in its linearized approximation), rendering the governing equations formally identical to those of Stokes flow in viscous fluids.
After deriving a general solution to the problem, an isoperimetric constraint is introduced at the boundary to enforce inextensibility. This effect can be physically realized, for example, by attaching an inextensible elastic rod with negligible bending stiffness to the perimeter. Although the combined imposition of material incompressibility and boundary inextensibility theoretically prevents any deformation of the disc, it is shown that the problem still admits non-trivial solutions. This apparent paradox is resolved by recognizing the approximations inherent in the linearized theory, as confirmed by a geometrically nonlinear numerical analysis.
Nonetheless, the linear solution retains significance: it may represent a valid stress distribution within a rigid system and can identify critical conditions of interest for design applications. 
\end{abstract}

\paragraph{Keywords}
Incompressible linear elasticity \textperiodcentered\
Isoperimetric constraint \textperiodcentered\
Complex variable solution \textperiodcentered\
Bifurcation

\section{Introduction}
\label{sec:introduction}

Linear elastic solutions for circular discs subjected to various loading conditions are fundamental to numerous engineering applications and have a long-standing history in mechanics, dating back to the classical Hertz contact problem. Professor Selvadurai, whose memory this article is dedicated to, made significant contributions to this area by deriving several analytical solutions for circular discs acting as inclusions (see, among his other works, \cite{selvadurai1, selvadurai2, selvadurai3}). 

This article addresses the problem of a linear elastic disc under general loading conditions, assuming that the material is incompressible. This constraint is enforced in the linearized framework by imposing that the trace of the strain tensor vanishes\footnote{
In the context of finite deformations, the isochoric constraint requires the deformation gradient to have unit determinant. This aspect is discussed in detail in Section \ref{vincolazzi}.
}. This assumption leads to a set of governing equations analogous to those describing the slow viscous flow of a Newtonian fluid. The solution is first obtained using complex variable potentials, adapted to enforce the isochoric (volume-preserving) condition, and some representative loading cases are examined. In the second part of the article, an isoperimetric constraint is imposed on the boundary of the disc. The simultaneous enforcement of both incompressibility and the isoperimetric condition in a two-dimensional setting might be expected to preclude any solution involving non-zero strain within the disc. However, this belief is shown to be incorrect: solutions for the \lq doubly constrained disc' are indeed found, provided certain conditions are met. 

On the other hand, evidence that the \lq doubly constrained disc' must behave as a rigid body is provided by a nonlinear finite element simulation of an incompressible disc coated with a nearly inextensible elastic rod carried out under the plane strain assumption. The simulation results highlight a fundamental inconsistency between linear and nonlinear elasticity theories. Similarly, a linear bifurcation analysis of an incompressible disc with an axially inextensible coating predicts a finite, rather than infinite, buckling load \cite{gaibotti2024bifurcations}. 
Similar inconsistencies have been previously observed in the context of anisotropic elasticity for elastic discs, especially in the works of Fosdick, Freddi, and Royer-Carfagni \cite{royer_1, royer_2, royer_3}. Nonetheless, we argue that the linearized solution presented here retains significant value, both theoretically and in practical applications related to engineering design.

From a theoretical point of view, the existence of a linearized solution for stress and strain in an elastic system that, due to applied constraints, should behave as a rigid body may be consistent with the attempt to define a state of stress for a rigid body as the limit of a sequence of elastic states \cite{grioli, royer_10, royer_11}.
From a practical perspective, we emphasize that the linearized solution remains a valuable tool for design purposes, as real-world constraints are never perfectly rigid. We refer to this practical consideration as the \lq bathyscaphe lesson' for reasons that will be clarified later.

The article is organized as follows. In Section~\ref{sec:problemform}, the incompressible elastic disc is solved using complex potentials, following the introduction of the incompressibility constraint in Section~\ref{sec:incomp}. Section~\ref{sec:soluzioni} presents examples of solutions for the incompressible elastic disc, while Section~\ref{vincolazzi} concludes the article with an analysis of the disc, in which both in-plane incompressibility and boundary isoperimetricity are simultaneously enforced.

\section{Incompressible linear elasticity}
\label{sec:incomp}

\paragraph{In a three-dimensional context} incompressibility of a material under linearized kinematics requires the displacement field $\bu$ to be solenoidal,
\begin{equation}
\label{solenoidu}
    \diver \bu = 0,
\end{equation}
which corresponds to $\tr \bvarepsilon = 0$, where $\bvarepsilon = (\nabla \bu + \nabla \bu^T)/2$ is the strain tensor. The constitutive equations of incompressible and linear elasticity relate the stress tensor $\bsigma$ to the purely deviatoric strain tensor as
\begin{equation}
\label{stresses}
    \bsigma = - p\, \bI + 2 \mu\, \bvarepsilon,
\end{equation}
where $\bI$ is the unit tensor, $\mu$ is the shear modulus, while 
\begin{equation}
    p = -\frac{1}{3} \tr \bsigma, 
\end{equation}
is the mean stress with reversed sign, or the \lq pressure', which remains undetermined from the constitutive equations and plays the role of a Lagrangian multiplier, to be added to the strain energy $W = \mu \, \bvarepsilon \cdot \bvarepsilon$ to obtain the stress as
\begin{equation}
    \bsigma = \frac{\partial }{\partial \bvarepsilon} \left( -p \tr \bvarepsilon + W \right).
\end{equation}
In the absence of body forces, the equilibrium equation,
\begin{equation}
    \diver \bsigma = \bzero , 
\end{equation}
leads to the Navier equation for the displacement field $\bu$
\begin{equation}
\label{navier}
    \mu\, \nabla^2 \bu - \nabla p = 0,
\end{equation}
where $\nabla^{2}$ is the Laplacian operator.

Incompressibility condition of eq.~\eqref{solenoidu}, and Navier equation \eqref{navier} show that $p$ is a harmonic function, i.e.,
\begin{equation}
    \nabla^2 p = 0.
\end{equation}

\paragraph{In a two-dimensional context} the treatment closely follows that for Stokes flow, see \cite{milne}. 
For an incompressible material, a {\it stream function} $\psi$ can be introduced to define the components of the displacement vector $\bu$ as
\begin{equation}
\label{stream}
    u_1 = - \psi_{,2}, \qquad
    u_2 = \psi_{,1}, 
\end{equation}
in which a comma denotes differentiation, so that $\psi_{,i}=\partial{\psi}/\partial x_{i}$. Using the Schwarz theorem (the equality of mixed partial derivatives), the incompressibility constraint is automatically satisfied,
\begin{equation}
    u_{1,1} + u_{2,2} = - \frac{\partial^2 \psi(x_1,x_2)}{\partial x_1 \partial x_2} +
    \frac{\partial^2 \psi(x_1,x_2)}{\partial x_2 \partial x_1} = 0.
\end{equation}
The Navier equation (\ref{navier}) leads to
\begin{equation} 
\label{biharm}
    \psi_{,1111} + 2 \psi_{,1122} + \psi_{,2222} = 0,
\end{equation}
which is the biharmonic equation that also governs slow viscous flow of a fluid, or \lq Stokes flow'. The latter equation has to be complemented with 
\begin{equation}
\label{harmonic}
    p_{,11} + p_{,22} = 0.
\end{equation}

\section{The incompressible elastic circular disc}
\label{sec:problemform}

\begin{figure}[hbt!]
	\centering
    \includegraphics[keepaspectratio, scale=1.1]{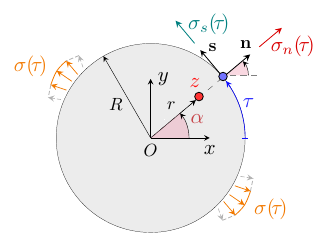} 
	\caption{The elastic disc subject to an assigned self-equilibrated traction distribution $\sigma(\tau)$ on the boundary. The length of the circular arc that connects the boundary point  $\tau$ and $(R,0)$ is $\alpha R$. Note the cylindrical $\alpha, r$ and Cartesian $x,y$ coordinates, and the outward normal $\bn$ and tangent $\bs$ unit vectors.}
	\label{fig:incomp_disk}
\end{figure}

Consider a circular disc centered at the origin of an $x$--$y$ Cartesian coordinate system, as shown in Fig.~\ref{fig:incomp_disk}. The disc consists of an incompressible elastic material, and a self-equilibrated load distribution acts on its boundary. In this section, we evaluate the elastic fields inside the disc and on its boundary for every possible load distribution.
%

To solve the problem, the so-called Wirtinger calculus is used \cite{koor2023shorttutorialwirtingercalculus}. Two independent complex variables, $z$ and $\overline{z}$, are defined in terms of Cartesian coordinates as $z = x + iy$ and $\overline{z} = x - iy$, where the bar denotes complex conjugation, and $x$ and $y$ represent the coordinates of the point. The partial derivatives in Cartesian coordinates can be expressed via the Wirtinger derivatives as
\begin{equation}
\label{Wderiv}
    \frac{\partial}{\partial x} = \frac{\partial}{\partial z} + \frac{\partial}{\partial\overline{z}}, 
    \quad 
    \frac{\partial}{\partial y} = -\frac{1}{i}\left(\frac{\partial}{\partial z} - \frac{\partial}{\partial\overline{z}}\right).
\end{equation}

For simplicity, let us first assume that the disc has a unit radius, $R = 1$. Having introduced complex variables, any point on the boundary of the disc, which was denoted by $\tau$, is defined as follows:
\begin{equation}
\label{marianna}
    \tau = \exp(i\alpha) = \cos\alpha + i\sin\alpha, 
    \quad 
    \frac{1}{\tau} = \exp(-i\alpha) = \cos\alpha - i\sin\alpha. 
\end{equation}
where $\alpha$ is the angle between the $x$--axis and the normal to the boundary of the disc at point $\tau$, Fig.~\ref{fig:incomp_disk}.

The stream function $\psi$ is biharmonic, eq.~\eqref{biharm}, and therefore it can be expressed using the following Goursat representation (similar to that used in \cite{brubeck2022lightning}):
\begin{equation}
\label{psi_Refg}
    \psi = \Rp \left[\overline{z} f(z) + g(z)\right],
\end{equation}
where $f(z)$, $g(z)$ are arbitrary holomorphic functions, satisfying 
\begin{equation}
    \frac{\partial f}{\partial\overline{z}} = 0, 
    \quad 
    \frac{\partial g}{\partial\overline{z}} = 0, 
    \quad 
    \frac{\partial\overline{f}}{\partial z} = 0, 
    \quad 
    \frac{\partial\overline{g}}{\partial z} = 0.
\end{equation}

Inside the circle $\left|z\right|<1$, holomorphic functions can be represented through complex Taylor series with constant complex coefficients $A_{k}$ and $B_k$ as 
\begin{equation}
\label{fg}
    f(z) = \sum^{\infty}_{k=0} A_{k}z^{k}, 
    \quad 
    g(z)=\sum^{\infty}_{k=0} B_{k}z^{k},
\end{equation}
where 
\begin{equation}
    A_{k} = \frac{f^{(k)}(0)}{k!}, 
    \quad 
    B_{k} = \frac{g^{(k)}(0)}{k!}.
\end{equation}

A substitution of expressions \eqref{fg} into \eqref{psi_Refg} leads to 
\begin{equation}
\label{psi}
    \psi = \frac{1}{2} \sum_{k=0}^{\infty} \Big(A_k \overline{z} z^k + \overline{A}_k z \overline{z}^k + 
    B_k z^k + \overline{B}_k \overline{z}^k \Big).
\end{equation}

From eqs.~\eqref{stream} and \eqref{Wderiv}, the displacement components become
\begin{equation}
\label{u}
    u_{1} = -\frac{\partial\psi}{\partial y} = \frac{1}{i} \left(\frac{\partial\psi}{\partial z} - 
    \frac{\partial\psi}{\partial\overline{z}}\right), 
    \quad 
    u_{2} = \frac{\partial\psi}{\partial x} = \frac{\partial\psi}{\partial z} + 
    \frac{\partial\psi}{\partial\overline{z}}.
\end{equation}

Thus, using eqs.~\eqref{Wderiv}--\eqref{u}, the displacement components can be obtained in terms of series as follows:
\begin{equation}
\label{comp}
    \begin{aligned}
    &u_1(z) = \frac{1}{2i} \sum_{k=1}^{\infty} k \Big[z^{k-1}\Big(\overline{z}A_k+B_k\Big)-\overline{z}^{k-1}\Big(z\overline{A}_k+\overline{B}_k\Big)\Big] + 
    \frac{1}{2i} \sum_{k=0}^{\infty} \Big(\overline{z}^k\, \overline{A}_k-z^kA_k\Big), \\[3mm] 
    &u_2(z) = \frac{1}{2} \sum_{k=1}^{\infty} k \Big[z^{k-1}\Big(\overline{z}A_k+B_k\Big)+\overline{z}^{k-1}\Big(z\overline{A}_k+\overline{B}_k\Big)\Big] + 
    \frac{1}{2} \sum_{k=0}^{\infty} \Big(\overline{z}^k \overline{A}_k+z^kA_k\Big).
\end{aligned}
\end{equation}

Combining the two above expressions, the following representation for the complex displacement is obtained
\begin{equation}
\label{9}
    u = u_{1} + i\, u_{2} = i \sum^{\infty}_{k=0} A_{k}z^{k} + 
    i\sum^{\infty}_{k=1}k\overline{z}^{k-1}\Big(z\overline{A}_{k}+\overline{B}_{k} \Big). 
\end{equation}

A possible rigid-body displacement can be eliminated by requiring the two constraints 
\begin{equation}
    u(0)=0, \quad u_{2}(1)=0, 
\end{equation}
so that, from eqs.~\eqref{comp}, \eqref{9}, one obtains
\begin{equation}
\label{A)}
    A_{0} + \overline{B}_{1} = 0,
\end{equation}
and
\begin{equation}
\label{reAB}
    \sum^{\infty}_{k=1} (k+1) \Rp(A_{k}) + \sum^{\infty}_{k=2} k \Rp(B_{k}) = 0.
\end{equation}

The Navier equation \eqref{navier} can be written in component form as
\begin{equation}
\label{eq:navier_components}
    \mu \nabla^{2} u_{1} - p_{,1} = 0, 
    \quad 
    \mu \nabla^{2} u_{2} - p_{,2} = 0,  
\end{equation}
in which $p$, the Lagrangian multiplier associated with the incompressibility constraint, is a harmonic function, eq.~\eqref{harmonic}. As such, it can be represented by the real part of a holomorphic function, which, inside the circle, can be represented through the series
\begin{equation}
\label{p}
    p(z)=
    \frac{1}{2}\, \sum^{\infty}_{k=0} C_{k}z^{k} + 
    \frac{1}{2}\, \sum^{\infty}_{k=0} \overline{C}_{k}\overline{z}^{k},
\end{equation}
where $C_{k}$ are unknown complex coefficients.

The derivatives of $p(z)$ can be obtained from eq.~\eqref{p} by using eq.~\eqref{Wderiv}, as
\begin{equation}
\label{derivp}
    p_{,1}(z) = \frac{1}{2} \sum_{k=1}^{\infty} k\Big( C_k  z^{k-1} + \overline{C}_k\overline{z}^{k-1}\Big), 
    \quad 
    p_{,2}(z) = \frac{i}{2} \sum_{k=1}^{\infty} k\Big( C_k  z^{k-1} - \overline{C}_k\overline{z}^{k-1}\Big).
\end{equation}

The form of the Laplacian operator
\begin{equation}
    \nabla^{2} = 4\frac{\partial^{2}}{\partial z\partial\overline{z}},   
\end{equation}
implies 
\begin{equation}
\label{Laplacian}
    \nabla^{2}u_{1} = 4\frac{\partial^{2}u_1}{\partial z\partial\overline{z}}, 
    \quad 
    \nabla^{2}u_{2} = 4\frac{\partial^{2}u_2}{\partial z\partial\overline{z}}.    
\end{equation}

A substitution of eqs.~\eqref{comp} in eq.~\eqref{Laplacian} yields
\begin{equation}
\label{nablau}
    \begin{aligned}
    &\nabla^{2}u_1 = -2i \sum_{k=2}^{\infty} k(k-1)\Big(A_k z^{k-2} - 
    \overline{A}_k \overline{z}^{k-2}\Big), \\[3mm] 
    &\nabla^{2}u_2 = 2 \sum_{k=2}^{\infty} k(k-1)\Big(A_k z^{k-2} + 
    \overline{A}_k \overline{z}^{k-2}\Big).
    \end{aligned}
\end{equation}

Eqs.~\eqref{derivp} and \eqref{nablau} substituted into eqs.~\eqref{eq:navier_components} yield a new form for the Navier equations
\begin{equation}
\begin{aligned}
    &\sum_{k=2}^{\infty} k(k-1)\Big(\overline{A}_k \overline{z}^{k-2}-A_k z^{k-2}\Big) = 
    -\frac{i}{4\mu} \sum_{k=1}^{\infty} k\Big(C_k z^{k-1} + \overline{C}_k \overline{z}^{k-1}\Big), \\[3mm]
    &\sum_{k=2}^{\infty} k(k-1)\Big(\overline{A}_k \overline{z}^{k-2}+A_k z^{k-2}\Big) = 
    \frac{i}{4\mu} \sum_{k=1}^{\infty} k\Big(C_k z^{k-1} - \overline{C}_k \overline{z}^{k-1}\Big),
\end{aligned}
\end{equation}

From the latter equations, we get
\begin{equation}
    \sum^{\infty}_{k=2} A_{k}k\left(k-1\right)z^{k-2} = 
    \frac{i}{4\mu} \sum^{\infty}_{k=1} C_{k}kz^{k-1},
\end{equation}
and
\begin{equation}
\label{C}
    C_{k} = -4\mu i(k+1)A_{k+1}, \quad k \geq 1.
\end{equation}

The tractions prescribed on the boundary of the disc $\tau\overline{\tau}=1$ are assumed to be represented through the following complex Fourier series
\begin{equation}
\label{expanssigma}
    \sigma\left(\tau\right) = \sigma_{n}\left(\tau\right) + i\sigma_{s}\left(\tau\right) = 
    \sum^{\infty}_{k=1} D_{-k}\frac{1}{\tau^{k}} + \sum^{\infty}_{k=0} D_{k}\tau^{k},   
\end{equation}
with $\sigma_{n}(\tau)$, $\sigma_{s}(\tau)$ being the normal and shear components of the tractions, Fig.~\ref{fig:incomp_disk}. Note that consideration of equilibrium allows one to obtain the restrictions \cite{mogilevskaya2001galerkin}
\begin{equation}
\label{equilib}
    D_{-1} = 0, \quad \Ip D_{0}=0.
\end{equation}

The representation of the boundary tractions in Cartesian coordinates can be written as  \cite{linkov1998complex} 
\begin{equation}
\label{complextrac}
    \sigma\left(\tau\right) = \exp(-i\alpha)\, t(\tau),
\end{equation}
in which $t(\tau) = t_{1}(\tau) + i\, t_{2}(\tau)$ is the complex representation for tractions in the Cartesian coordinate system.

Use of the standard expressions for the tractions in terms of stresses and the expressions for the components of the outward vector normal to the circle leads to (details are omitted for brevity)
\begin{equation}
\label{t}
    t_{1} = \sigma_{11}\cos\alpha + \sigma_{12}\sin\alpha, 
    \qquad
    t_{2} = \sigma_{12}\cos\alpha + \sigma_{22}\sin\alpha, 
\end{equation}
while rewriting eq.~\eqref{stresses} as 
\begin{equation}
\label{sigij}
    \sigma_{ij} = -p\, \delta_{ij} + 2\mu\left(\varepsilon_{ij} - 
    \frac{1}{3}\delta_{ij}\varepsilon_{kk}\right), 
\end{equation}
and noticing that $\varepsilon_{kk}=0$, eqs.~\eqref{t} become 
\begin{equation}
\label{tracomp}
    t_{1} = (-p+2\mu\varepsilon_{11})\cos\alpha+2\mu\varepsilon_{12}\sin\alpha, 
    \quad
    t_{2} = 2\mu\varepsilon_{12}\cos\alpha+(-p+2\mu\varepsilon_{22})\sin\alpha.
\end{equation}

The expressions for the strains in terms of displacements can be obtained using the differentiation rules, eq.~\eqref{Wderiv}, as
\begin{equation}
\label{strains}
    \begin{gathered}
    \varepsilon_{11} = \frac{\partial u_{1}}{\partial x}=\frac{\partial u_{1}}{\partial z}+\frac{\partial u_{1}}{\partial\overline{z}}, 
    \quad
    \varepsilon_{22} = \frac{\partial u_{2}}{\partial y}=i\left(\frac{\partial u_{2}}{\partial z}-\frac{\partial u_{2}}{\partial\overline{z}}\right), \\[3mm]
    \varepsilon_{12} = \frac{1}{2}\left(\frac{\partial u_{1}}{\partial y}+\frac{\partial u_{2}}{\partial x}\right)=\frac{1}{2}\left[i\left(\frac{\partial u_{1}}{\partial z}-\frac{\partial u_{1}}{\partial\overline{z}}\right)+\left(\frac{\partial u_{2}}{\partial z}+\frac{\partial u_{2}}{\partial\overline{z}}\right)\right].
    \end{gathered}
\end{equation}

From eqs.~\eqref{comp}, the following representations can be obtained
\begin{equation}
\label{partialu}
    \begin{aligned}
    &\frac{\partial u_{1}}{\partial z} = \frac{1}{2i}\sum_{k=2}^{\infty}k(k-1)z^{k-2}\Big(\overline{z}A_k+B_k\Big)-\frac{1}{2i}\sum_{k=1}^{\infty}k\Big(A_k z^{k-1}+\overline{A}_k \overline{z}^{k-1}\Big) ,\\[3mm]
    &\frac{\partial u_{1}}{\partial \overline{z}} = \frac{1}{2i}\sum_{k=1}^{\infty}k\Big(A_k z^{k-1}+\overline{A}_k \overline{z}^{k-1}\Big)-\frac{1}{2i}\sum_{k=2}^{\infty}k(k-1)\overline{z}^{k-2}\Big(z\overline{A}_k+\overline{B}_k\Big), \\[3mm]
    &\frac{\partial u_2}{\partial z} = \frac{1}{2}\sum_{k=2}^{\infty}k(k-1)z^{k-2}\Big(\overline{z}A_k+B_k\Big)+\frac{1}{2}\sum_{k=1}^{\infty}k\Big(A_k z^{k-1}+\overline{A}\overline{z}^{k-1}\Big) , \\[3mm]
    &\frac{\partial u_2}{\partial \overline{z}} = \frac{1}{2}\sum_{k=1}^{\infty}k\Big(A_k z^{k-1}+\overline{A}_k \overline{z}^{k-1}\Big)+\frac{1}{2}\sum_{k=1}^{\infty}k(k-1)\overline{z}^{k-2}\Big(z\overline{A}_k + \overline{B}_k \Big),
    \end{aligned}
\end{equation}
which, substituted into eqs.~\eqref{strains}, yield
\begin{equation}
\label{straincom}
    \begin{aligned}
    &\varepsilon_{11} = -\varepsilon_{22} = 
    \frac{1}{2i} \sum_{k=2}^{\infty} k(k-1)\Big[z^{k-2}\Big(\overline{z}A_k+B_k\Big) - \overline{z}^{k-2}\Big(z\overline{A}_k+\overline{B}_k\Big)\Big], \\[3mm]
    &\varepsilon_{12} = \frac{1}{2} \sum_{k=2}^{\infty} k(k-1)\Big[z^{k-2}\Big(\overline{z}A_k + B_k\Big) + \overline{z}^{k-2}\Big(z\overline{A}_k+\overline{B}_k\Big)\Big].
    \end{aligned}
\end{equation}

It can be easily verified that the above equations satisfy the incompressibility constraint, $\varepsilon_{11}+\varepsilon_{22}=0$. 

Considering a point $\tau$ on the boundary, eq.~\eqref{marianna}, eqs.~\eqref{complextrac} and \eqref{tracomp} lead to the representation of the tractions at the boundary 
\begin{equation}
    t=\big[-p+\mu(\varepsilon_{11}+\varepsilon_{22})\big]\exp(i\alpha)+\mu\big(2i\epsilon_{12}+\varepsilon_{11}-\varepsilon_{22}\big)\exp(-i\alpha),
\end{equation}
or, from eq.~\eqref{complextrac} and accounting the incompressibility constraint $\varepsilon_{11}+\varepsilon_{22}=0$
\begin{equation}
\label{sigma}
    \begin{aligned}
    \sigma = -p - \mu(\varepsilon_{22}-\varepsilon_{11}-2i\varepsilon_{12}) \exp(-2i\alpha).
    \end{aligned}
\end{equation}

The use of eqs.~\eqref{p} and \eqref{straincom} with $z=\tau$ in eq.~\eqref{sigma}, with $\exp(-2i\alpha)=1/\tau^2$), leads to
\begin{equation}
\label{boundary sigma}
    \sigma = -\frac{1}{2} \sum_{k=0}^{\infty} \Big(C_k\tau^k+\overline{C}_k\frac{1}{\tau^k}\Big) + 
    2\mu i \sum_{k=2}^{\infty} k(k-1)\Big(\overline{A}_k\frac{1}{\tau^{k-1}}+\overline{B}_k\frac{1}{\tau^{k}}\Big).
\end{equation}

A substitution of the expansion of eq.~\eqref{expanssigma} with eq.~\eqref{C} in eq.~\eqref{boundary sigma} provides the expression
\begin{equation}
\label{DvsAB}
    \sum^{\infty}_{k=1} D_{-k} \frac{1}{\tau^{k}} + 
    \sum^{\infty}_{k=0} D_{k}\tau^{k} = 
    -\Rp(C_0) + 2{\mu}i \sum^{\infty}_{k=2} k\Big[A_k\tau^{k-1}+(k-2)\overline{A}_k\frac{1}{\tau^{k-1}}+(k-1)\overline{B}_k\frac{1}{\tau^k}\Big].
\end{equation}

The use of the orthogonality property of the Fourier series, eq.~\eqref{DvsAB} yields 
\begin{equation}
\label{coeff_incomp}
    \begin{gathered}
    D_0= -\Rp(C_0) , 
    \quad
    D_1=4i\mu A_2, \\[3mm]
    D_{k}=2i\mu A_{k+1}(k+1), \quad k \geq 1, \\[3mm]
    D_{-k}=2i\mu\left[\overline{A}_{k+1}(k^2-1)+k\overline{B}_{k}(k-1)\right], \quad k \geq 2, 
    \end{gathered}
\end{equation}
from which it follows that
\begin{equation}
\label{coeff_incomp2}
    \begin{gathered}
    \Rp(C_0)=-D_0, 
    \quad
    A_2=-\frac{i}{4\mu}D_1, \\[3mm]
    A_{k+1}=-\frac{i}{2\mu}\frac{D_{k}}{k+1}, 
    \quad
    \overline{B}_k=-\frac{i}{2\mu}\frac{D_{-k}+(k-1)\overline{D}_{k}}{k(k-1)}, \quad k \geq 2 .
    \end{gathered}
\end{equation}
Eqs.~\eqref{coeff_incomp2}$_{2-3}$ can be combined together, resulting in
\begin{equation}
\label{Ak1}
    A_{k+1} = -\frac{i}{2\mu}\frac{D_{k}}{k+1}, \quad k \geq 1,
\end{equation}
and eqs.~\eqref{reAB} and \eqref{coeff_incomp2} lead to 
\begin{equation}
\label{reA1}
    2\Rp(A_1) = -\frac{1}{2\mu}\left\{\frac{3}{2}\Ip(D_1) + 
    \sum_{k=2}^{\infty} \Ip\left[\frac{(k+1)D_{-k}+(k-1)D_k}{k^2-1}\right]\right\}.
\end{equation}
The stream function $\psi(z)$ in eq.~\eqref{psi} can be written as
\begin{equation}
\label{psi_expl}
    \psi(z)=\Rp\Big[\overline{z}\big(A_0+\overline{B}_1\big)+B_0+\overline{z}zA_1\Big]+\frac{1}{2}\sum_{k=1}^{\infty}\Big(A_{k+1}\overline{z}z^{k+1}+\overline{A}_{k+1}z\overline{z}^{k+1}\Big)\frac{1}{2}+\sum_{k=2}^{\infty}\Big(B_k z^k+\overline{B}_k\Big)\overline{z}^k,
\end{equation}
where the substitution of eqs.~\eqref{A)}, \eqref{coeff_incomp2}$_4$ and \eqref{Ak1} into eq. \eqref{psi_expl} leads to
\begin{multline}
\label{psi_D}
    \psi(z)=\Rp\big(B_0\big)-\frac{3}{8\mu}\overline{z}z\Ip(D_1)-\frac{\overline{z}z}{4\mu}\sum_{k=2}^{\infty}\Ip\left[\frac{(k+1)D_{-k}+(k-1)D_k}{k^2-1}\right] \\[3mm]
    +\frac{i}{4\mu}\sum_{k=1}^{\infty}\frac{\overline{D}_k\,z\overline{z}^{k+1}-D_k\,\overline{z}z^{k+1}}{k+1}+\frac{i}{4\mu}\sum_{k=2}^{\infty}\frac{\big[\overline{D}_{-k}+(k-1)D_k\big]z^k-\big[D_{-k}+(k-1)\overline{D}_k\big]\overline{z}^{k}}{k(k-1)}.
\end{multline}

A substitution of eqs.~\eqref{coeff_incomp2} in eq.~\eqref{9} yields the following expression for the complex displacement 
\begin{multline}
\label{u_coeff}
    u(z)=i\left\{A_0+\overline{B}_1+2z \Rp A_1+\frac{i}{2\mu}\left[z\overline{z}\overline{D}_1-\sum_{k=1}^{\infty}\frac{D_k}{k+1}z^{k+1}+\sum_{k=2}^{\infty}\overline{D}_k\overline{z}^{k-1}(z\overline{z}-1)\right.\right. \\[3mm]
    \left.\left.-\sum_{k=2}^{\infty}\frac{D_{-k}}{k-1}\overline{z}^{k-1}\right]\right\}.
\end{multline}

After substituting eqs.~\eqref{A)} and \eqref{reA1} in eq.~\eqref{u_coeff}, a new expression for the complex displacement is obtained 
\begin{multline}
\label{u_D}
    2\mu\, u(z) = iz\left\{-\frac{3}{2}\Ip(D_1)-\sum_{k=2}^{\infty}\Ip\left[\frac{(k+1)D_{-k}+(k-1)D_{k}}{k^2-1}\right]\right\}-z\overline{z}\overline{D}_1+\sum_{k=2}^{\infty}\overline{D}_{k}\overline{z}^{k-1}(1-z\overline{z}) \\[3mm]
    +\sum_{k=2}^{\infty}\frac{D_{-k}}{k-1}\overline{z}^{k-1}+\sum_{k=1}^{\infty}\frac{D_{k}}{k+1}z^{k+1}.
\end{multline}
At the boundary of the disc, $z=\tau$ and $\tau\overline{\tau}=1$, the complex displacement assumes the form
\begin{multline}
\label{utau}
    2\mu\, u(\tau) = i\tau\left\{-\frac{3}{2}\Ip(D_1)-\sum_{k=2}^{\infty}\Ip\left[\frac{(k+1)D_{-k}+(k-1)D_{k}}{k^2-1}\right]\right\}-\overline{D}_1 \\[3mm]
    +\sum_{k=2}^{\infty}\frac{{D}_{-k}}{k-1}\frac{1}{\tau^{k-1}}+\sum_{k=1}^{\infty}\frac{{D}_{k}}{k+1}\tau^{k+1}.
\end{multline}
A substitution of eqs.~\eqref{coeff_incomp2}$_4$ and \eqref{Ak1} into \eqref{straincom} provides the following representation for the strains satisfying incompressibility $\varepsilon_{11}(z)+\varepsilon_{22}(z)=0$ inside the disc, in terms of the coefficients of eq.~\eqref{expanssigma},
\begin{equation}
\label{paolagatti}
    \varepsilon_{22}(z)-\varepsilon_{11}(z)-2i\varepsilon_{12}(z) = 
    \frac{1}{\mu}\left[z\overline{D}_1 + \sum_{k=2}^{\infty} k\overline{D}_k\overline{z}^{k-2}(z\overline{z}-1) + \sum_{k=2}^{\infty} \left(\overline{D}_k-D_{-k}\right)\overline{z}^{k-2}\right],
\end{equation}
a substitution of eqs.~\eqref{p} and \eqref{straincom} in eq.~\eqref{sigij} and use of eqs.~\eqref{coeff_incomp2} and \eqref{Ak1}, provides the corresponding stresses
\begin{equation}
\label{stresses_fields}
    \begin{gathered}
    p(z) = -\frac{\sigma_{11}(z)+\sigma_{22}(z)}{2} = -D_0-2\sum_{k=1}^{\infty}\Rp\Big(D_{k}z^k\Big), \\[3mm]
    \sigma_{22}(z)-\sigma_{11}(z)-2i\sigma_{12}(z) = 2\left[z\overline{D}_1 + 
    \sum_{k=2}^{\infty} k\overline{D}_k\overline{z}^{k-2}(z\overline{z}-1) + 
    \sum_{k=2}^{\infty} \left(\overline{D}_k-D_{-k}\right)\overline{z}^{k-2}\right],
    \end{gathered}
\end{equation}

Eqs.~\eqref{paolagatti} and \eqref{stresses_fields} represent the solution of the elastic disc of unit radius made of incompressible material. It can be shown in a straightforward manner that similar expressions for the elastic fields within the disc of radius $R$ are, in addition to the incompressibility constraint $\varepsilon_{11}(z)+\varepsilon_{22}(z)=0$, 

\begin{multline}
\label{u_D_R}
    \frac{2\mu\, u(z)}{R} = 
    i\frac{z}{R}\left\{-\frac{3}{2}\Ip(D_1) - \sum_{k=2}^{\infty} \Ip\left[\frac{(k+1)D_{-k}+(k-1)D_{k}}{k^2-1}\right]\right\} - \frac{z\overline{z}}{R^2}\overline{D}_1 \\[3mm]
    + \sum_{k=2}^{\infty} \overline{D}_{k}\frac{\overline{z}^{k-1}}{R^{k-1}}(1-\frac{z\overline{z}}{R^2}) 
    + \sum_{k=2}^{\infty} \frac{D_{-k}}{k-1}\frac{\overline{z}^{k-1}}{R^{k-1}}+\sum_{k=1}^{\infty}\frac{D_{k}}{k+1}\frac{z^{k+1}}{R^{k+1}}.
\end{multline}

\begin{equation}
\label{uepsR}
    \varepsilon_{22}(z)-\varepsilon_{11}(z)-2i\varepsilon_{12}(z)= \frac{1}{\mu}\left[\frac{z}{R}\overline{D}_1+\sum_{k=2}^{\infty}k\overline{D}_k\frac{\overline{z}^{k-2}}{R^{k-2}}(\frac{z\overline{z}}{R^2}-1)+\sum_{k=2}^{\infty}\left(\overline{D}_k-D_{-k}\right)\frac{\overline{z}^{k-2}}{R^{k-2}}\right],
\end{equation}

\begin{equation}
\label{stresses_fieldsR}
    \begin{gathered}
    p(z)=-\frac{\sigma_{11}(z)+\sigma_{22}(z)}{2}=-D_0-2\sum_{k=1}^{\infty}\Rp\Big(D_{k}\frac{z^k }{R^k}\Big), \\[3mm]
    \sigma_{22}(z)-\sigma_{11}(z)-2i\sigma_{12}(z)=2\left[\frac{z}{R}\overline{D}_1+\sum_{k=2}^{\infty}k\overline{D}_k\frac{\overline{z}^{k-2}}{R^{k-2}}(\frac{z\overline{z}}{R^2}-1)+\sum_{k=2}^{\infty}\left(\overline{D}_k-D_{-k}\right)\frac{\overline{z}^{k-2}}{R^{k-2}}\right],
    \end{gathered}
\end{equation}
where the prescribed load at the boundary, $z=\tau$ and $\tau\overline{\tau}=R^2$, can be expanded as
\begin{equation}
\label{expanssigmaR}
    \sigma(\tau) = \sum^{\infty}_{k=1} D_{-k}\frac{R^k}{\tau^{k}} + 
    \sum^{\infty}_{k=0} D_{k}\frac{\tau^{k}}{R^k}.   
\end{equation}

The expressions for the remaining elastic fields at the boundary are 
\begin{equation}
\label{fields_tau}
    \begin{gathered}
    \begin{multlined}
    2\mu\, u(\tau) = i\tau\left\{-\frac{3}{2}\Ip(D_1)-\sum_{k=2}^{\infty}\Ip\left[\frac{(k+1)D_{-k}+(k-1)D_{k}}{k^2-1}\right]\right\}-R\overline{D}_1 \\[3mm] 
    +\sum_{k=2}^{\infty}\frac{{D}_{-k}}{k-1}\frac{R^k}{\tau^{k-1}}+\sum_{k=1}^{\infty}\frac{{D}_{k}}{k+1}\frac{\tau^{k+1}}{R^k}, 
    \end{multlined} \\[3mm]
    \varepsilon_{22}(\tau)-\varepsilon_{11}(\tau)-2i\varepsilon_{12}(\tau)= \frac{1}{\mu}\left[\frac{z}{R}\overline{D}_1+\sum_{k=2}^{\infty}\left(\overline{D}_k-D_{-k}\right)\frac{R^{k-2}}{\tau^{k-2}}\right], \\[3mm]
    p(\tau)=-\frac{\sigma_{11}(\tau)+\sigma_{22}(\tau)}{2}=-D_0-2\sum_{k=1}^{\infty}\Rp\Big(D_{k}\frac{\tau^k }{R^k}\Big), \\[3mm]
    \sigma_{22}(\tau)-\sigma_{11}(\tau)-2i\sigma_{12}(\tau)=2\left[\frac{\tau}{R}\overline{D}_1+\sum_{k=2}^{\infty}\left(\overline{D}_k-D_{-k}\right)\frac{R^{k-2}}{\tau^{k-2}}\right].
    \end{gathered}
\end{equation}

\subsection{\texorpdfstring{Connection with the plane strain compressible case for $\nu \to 1/2$}{}}
\label{mantegazza}

The plane strain case of compressible elasticity treated in \cite{gaibotti2023elastic} is now considered. It is shown below that, in the isochoric limit, the solution of that problem $\nu \to 1/2$, coincides exactly with the solution obtained above directly for an incompressible material. This shows that the linear elastic solution for compressible materials converges smoothly to the incompressible case. 

The disc material is assumed to be isotropic and compressible, characterized by the shear modulus $\mu$ and Poisson's ratio $\nu$. As shown in eq.~(29) of \cite{mogilevskaya2008multiple}, and reproduced as eq.~(27) in \cite{gaibotti2023elastic}, the coefficients of the Fourier series expansions for the displacements and tractions at the boundary of the disc are interrelated (in the present notation) as
\begin{equation}
\label{D}
    \begin{gathered}
    D_{-1} = 0, \quad
    \frac{\kappa-1}{2\mu}D_{0} = 2\Rp\frac{H_{1}}{R}, \\[3mm]
    \frac{1}{2\mu}D_{-k} = \left(k-1\right)\frac{H_{1-k}}{R}, \quad k \geq 2, \\[3mm]
    \frac{\kappa}{2\mu}D_{k} = \left(k+1\right)\frac{H_{k+1}}{R}, \quad k \geq 1,
    \end{gathered}
\end{equation}
in which $\kappa=3-4\nu=1$. Moreover, the tractions at the boundary are given by eq.~\eqref{expanssigmaR} and the displacements there are represented as 
\begin{equation}
\label{albino}
    u(\tau) = u_{1}(\tau) + iu_{2}(\tau) = 
    \sum^{\infty}_{k=1} H_{-k}\frac{R^k}{\tau^k} + \sum^{\infty}_{k=0} H_{k}\frac{\tau^{k}}{R^k}.
\end{equation}
Assuming now that the displacements at the boundary satisfy the inextensibility constraint, eq.~\eqref{inexten}, eq.~\eqref{albino} yields (see eqs.~(34)--(36) of \cite{gaibotti2023elastic}),
\begin{equation}
\label{H}
    \Rp H_{1} = 0, \quad
    H_{2} = 0, \quad
    H_{k+1}	= \frac{k-1}{k+1}\overline{H}_{1-k}, \quad k \geq 2.
\end{equation}

An analysis of eqs.~\eqref{D}, \eqref{expanssigmaR}, and \eqref{H} leads to the conclusion that the load must have the form expressed by eq.~\eqref{admis}. The displacements and stresses within the disc can be found from the Kolosov-Muskhelishvili formulae \cite{muskhelishvili2013some} as
\begin{equation}
    \begin{gathered}
    2\mu\, u(z) = \kappa\varphi\left(z\right) - z\overline{\varphi'}\left(z\right) - \overline{\Psi}\left(z\right), \\[3mm]
    \sigma_{11}(z)+\sigma_{22}(z) = 4 \Rp \varphi'\left(z\right), \\[3mm]
    \sigma_{22}(z)-\sigma_{11}(z)-2i\sigma_{12}(z)	=2\left[z\overline{\varphi''}\left(z\right)+\overline{\Psi'}\left(z\right)\right].
    \end{gathered}    
\end{equation}
in which $\Psi$ should not be confused with the stream function used in the previous sections and the potentials were obtained in terms of displacement coefficients in \cite{mogilevskaya2008multiple}, see eq.~(51) of that article. 

Here, assuming $\nu=1/2$, eqs.~\eqref{D} and \eqref{H} can be rewritten in terms of the traction coefficients defined by eq.~\eqref{expanssigmaR} as
\begin{equation}
\label{eq:potentials}
    \varphi(z) = \frac{D_0}{2}z+\sum^{\infty}_{k=2}\frac{D_k}{k+1}\frac{z^{k+1}}{R^k}, 
    \quad
    \Psi(z) = -\sum^{\infty}_{k=2}D_k\frac{k}{k-1}\frac{z^{k-1}}{R^{k-2}}, 
\end{equation}  
so that the fields within the disc are
\begin{equation}
\label{expfields}
    \begin{gathered}
    \frac{2\mu\, u(z)}{R}	= \sum^{\infty}_{k=2}\frac{D_k}{k+1}\frac{z^{k+1}}{R^{k+1}}+\sum^{\infty}_{k=2}\overline{D_k}\frac{{\overline{z}^{k-1}}}{R^{k-1}}\left(\frac{k}{k-1}-\frac{z\overline{z}}{R^2}\right), \\[3mm]
    \sigma_{11}(z)+\sigma_{22}(z) = 2\Rp{D_0} + 4 \Rp \sum^{\infty}_{k=2}D_k\frac{z^{k}}{R^k}, \\[3mm]
    \sigma_{22}(z)-\sigma_{11}(z)-2i\sigma_{12}(z) = 2\sum^{\infty}_{k=2}k\overline{D}_k\frac{{\overline{z}^{k-2}}}{R^{k-2}}\left(\frac{z\overline{z}}{R^2}-1\right).
    \end{gathered}    
\end{equation}

A comparison between eqs.~\eqref{expfields} and eqs.~\eqref{finfieldsR}$_{1,4,5}$ suggests that the expressions for the stresses are the same in both equations, but the expressions for the complex displacement, eq.~\eqref{expfields}, differ in the term corresponding to a rigid-body displacement. The latter term can be added to the first line of eq.~\eqref{expfields} by assuming $u_2(R)=0$. Therefore, it can be concluded that, for the case of a prescribed admissible load at the disc boundary, the plane strain solution with $\nu=1/2$ is identical to that obtained in Section \ref{sec:problemform}.

Finally, the strains inside the disc can be obtained by using the following relations in the limit $\nu=1/2$ 
\begin{equation}
\label{expstrains}
    \begin{gathered}
    \varepsilon_{11}(z)+\varepsilon_{22}(z) = \frac{1-2\nu}{2\mu}[\sigma_{11}(z)+\sigma_{22}(z)]=0, \\[3mm]
    \varepsilon_{22}(z)-\varepsilon_{11}(z)-2i\varepsilon_{12}(z) = \frac{1}{2\mu}[\sigma_{22}(z)-\sigma_{11}(z)-2i\sigma_{12}(z)] = \sum^{\infty}_{k=2}k\frac{\overline{D}_k}{\mu}\frac{{\overline{z}^{k-2}}}{R^{k-2}}(\frac{z\overline{z}}{R^2}-1),
\end{gathered} 
\end{equation}
from which it can be concluded that all components of the strain tensor at the boundary of the disc vanish, as in the case studied in Section \ref{sec:problemform}, see the comment after eq.~\eqref{finfieldsR}.

\section{Examples of solutions for the incompressible elastic disc}
\label{sec:soluzioni}

A circular disc made of incompressible elastic material is considered, subject to a symmetric boundary traction distribution. 

As a first example, the validity of the approach to incompressible elasticity is checked through a trivial example, in which a uniform radial traction $\sigma(\tau)=-\Pi\,\bn$ is prescribed on the entire boundary of the disc ($\Pi$ is positive for compressive tractions). In this case, the only non-vanishing coefficient in eq.~\eqref{expanssigmaR} is $D_0$ and the treatment developed in the previous Section, eqs.~\eqref{stresses_fieldsR}, gives the expected trivial solution in which only $p(z)$ is different from zero and equal to the prescribed radial traction 
\begin{equation}
\label{p_hydro}
    p(z) = -D_0, \quad D_0 = -\Pi,
\end{equation}
while all displacements are null.

\begin{figure}[htb!]
    \centering
    \includegraphics[scale=1.10,keepaspectratio]{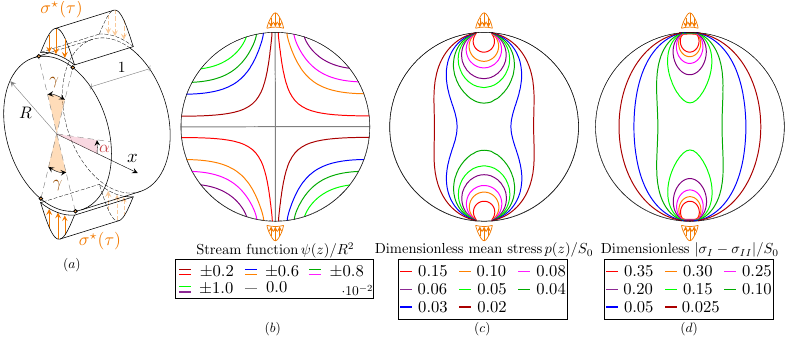}
    \caption{(a) The elastic disc made of incompressible material subjected to assigned non-uniform sinusoidal traction distribution $\sigma^{\star}(\tau)$ applied on the two small circular arcs $\alpha \in [\pm(\pi-\gamma)/2,\pm(\pi+\gamma)/2]$ centred at the angles $\pm\pi/2$. Contour lines of:  (b) the dimensionless stream function $\psi(z)/R^2$; (c) the dimensionless in-plane mean stress $p(z)/S_0$; (d) the dimensionless in-plane principal stress difference $|\sigma_{II}-\sigma_{II}|/S_0$.}
    \label{incomp_fields_plot}
\end{figure}

In a more elaborate example, a load acting on two symmetric portions of the boundary is applied. The traction is radial and varies along the two arcs spanned by the angle $\alpha$ varying within $[\pm(\pi-\gamma)/2,\pm(\pi+\gamma)/2]$, with the mid-points located at the angles $\alpha = \pm \pi/2$. The load is assumed with the shape of a sinusoidal function of $\alpha$ in the plane ($x,y$), Fig.~\ref{incomp_fields_plot}(a), as 
\begin{equation}
\label{sigma_sin}
    \sigma^{\star}(\alpha) = 
    \left\{
    \begin{array}{ll}
    -S_0\sin\Big[\frac{\pi}{\gamma}\Big(\alpha\mp \frac{\pi-\gamma}{2}\Big)\Big] & 
    \text{if } \alpha \in \big[\pm\frac{\pi-\gamma}{2}, \pm\frac{\pi+\gamma}{2}], \\[3mm]
    0 & 
    \text{elsewhere,}
    \end{array}
    \right.
\end{equation}
where $S_0$ is the traction amplitude.

The coefficients $D_{\pm k}$ involved in the series, eq.~\eqref{expanssigma}, can be computed using the same procedure reported in \cite{gaibotti2023elastic} and leading to
\begin{equation}
\label{int_D}
    \begin{aligned}
    &D_{\pm\pi/\gamma}=-\frac{S_0\gamma}{4\pi}e^{\frac{i\pi^2}{2\gamma}}\left(1+e^{\frac{i\pi^2}{\gamma}}\right), 
    \quad 
    \text{if } \frac{\pi}{\gamma} \in \mathbb{Z}, \\[3mm]
    &D_{m}=-\frac{S_0\,\gamma}{2\pi}e^{-\frac{1}{2}im(\pi+\gamma)}\frac{(1+e^{im\pi})(1+e^{im\gamma})}{m^2\gamma^2-\pi^2}.
    \end{aligned}
\end{equation}

Note that when $\sigma^{\star}(\alpha)$ becomes a constant function over the arc defined above, the expressions for the coefficients take the same form as those given in eq.~(48) of \cite{gaibottiprestressato}.

Under the assumed distribution of load, eq.~\eqref{sigma_sin}, Fig.~\ref{incomp_fields_plot} reports the contour plots of the dimensionless stream function $\psi(z)/R^2$, the in-plane dimensionless mean stress $p(z)/S_0$, and principal stress difference $|\sigma_{II}-\sigma_{I}|/S_0$  acting at every point $z$ within the disc, panels (b), (c), and (d), respectively. It should be noted that all symmetries are preserved and that the distributions of $p$ and $\sigma_I-\sigma_{II}$ are qualitatively similar, with a \lq bulb-like' shaped stress near the load application, where the stress state is more pronounced than in the outer regions.

\section{The problem of an incompressible circular disc enhanced with an isoperimetric constraint at its boundary}
\label{vincolazzi}

A circular disc made up of linear elastic (compressible) material was analyzed in \cite{gaibotti2023elastic, gaibotti2024bifurcations, gaibottiprestressato}, under the condition that its boundary be coated with an axially-inextensible Euler rod. In addition to effects related to bending moment and shear forces, the inextensibility of the rod enforces an {\it isoperimetric constraint} to the boundary of the disc. Therefore, if the material composing the disc is {\it both incompressible and deformed in plane strain}, the disc/coating system is expected not to deform when subject to any loading distribution, because otherwise one of the three constraints --inextensibility, or incompressibility, or plane strain deformation-- would be violated. In fact, the circle is the plane figure that encloses the maximum area for a given perimeter.

With the above premise, it is shown in this section that, surprisingly, when coated with an inextensible rod, the incompressible disc considered in the previous sections still admits plane strain solutions in which the strain field does not vanish. This is because both the inextensibility and incompressibility constraints are enforced in their linearized versions and thus are approximate. In particular, for a solid subjected to a displacement field $\bu$, the isochoric constraint and its linearization are  
\begin{equation}
    \det \Big(\bI + \nabla \bu\Big) = 1 
    \quad 
    \xRightarrow[\text{Linearization}]{} \quad \nabla \cdot \bu = 0,
\end{equation}
while, for an annular rod (in a cylindrical coordinate system) subjected to a displacement $\bu$, the axial inextensibility constraint  and its linearization are
\begin{equation}
\label{polpettone}
    \Big(\frac{\partial u_\alpha}{\partial \alpha}+u_r + R\Big)^2
    + \Big(\frac{\partial u_r}{\partial \alpha}-u_\alpha \Big)^2 = R^2, 
    \quad 
    \xRightarrow[\text{Linearization}]{} \quad \frac{\partial u_\alpha}{\partial \alpha}+u_r=0 .
\end{equation}
It is shown below that the approximation inherent to the linearization of the two constraints is sufficient to let the incompressible disc with isoperimetric restriction admit non-trivial solutions for self-equilibrated loads applied on its boundary.

\subsection{The disc with isoperimetric constraint}

It is assumed now that the displacements at the boundary of the disc of radius $R$ satisfy the isoperimetric constraint, without introducing a coating with an elastic rod possessing a bending stiffness. 

The inextensibility condition at the boundary $\tau\overline{\tau}=R^2$ is (see eq.~(31) in \cite{gaibotti2023elastic})
\begin{equation}
\label{inexten}
    \Rp \left(\frac{\partial u}{\partial\tau}\right) = 0. 
\end{equation} 

Eq.~\eqref{fields_tau}$_1$ allows to provide the derivative of the displacement as 
\begin{multline}
\label{dutau}
    \frac{\partial u}{\partial\tau} = 
    \frac{i}{2{\mu}}\left\{-\frac{3}{2} \Ip D_1 - 
    \sum_{k=2}^{\infty} \Ip\left[\frac{(k+1)D_{-k}+(k-1)D_{k}}{k^2-1}\right]\right\} \\[3mm]
    -\frac{1}{2\mu} \sum_{k=2}^{\infty}{D}_{-k} \frac{R^k}{\tau^{k}} +
    \frac{1}{2\mu} \sum_{k=1}^{\infty} {D}_{k}\frac{\tau^{k}}{R^k}, 
\end{multline}
in which the traction coefficients are defined by eq.~\eqref{expanssigmaR}. Therefore, the inextensibility constraint, eq.~\eqref{inexten}, can be rewritten as
\begin{equation}
\label{redutau}
    \begin{aligned}
    \Rp\Big(\frac{\partial u}{\partial\tau}\Big) = 
    -\frac{1}{2\mu} \sum_{k=2}^{\infty} \left[{{D}_{-k}}\frac{R^k}{\tau^{k}} + 
    \overline{{D}}_{-k}\frac{\tau^{k}}{R^k}\right] +
    \frac{1}{2\mu} \sum_{k=1}^{\infty} \left[{D}_{k}\frac{\tau^{k}}{R^k} + 
    \overline{D}_{k}\frac{R^k}{\tau^{k}}\right] = 0.
    \end{aligned}
\end{equation}
At this stage, the orthogonality conditions for the Fourier series on the boundary lead to the conclusion that the solution for the problem of a disc made of incompressible material and isoperimetrically constrained, eq.~\eqref{inexten}, is only possible when the following admissibility conditions are met for the  boundary  tractions: 
\begin{equation}
    D_1 = 0, 
    \qquad 
    D_{-k} = \overline{D}_{k}, \quad k \geq 2.
\end{equation}

As a consequence, any load at the boundary of the disc $\tau\overline{\tau}=R^2$ that provides a solution can be represented as 
\begin{equation}
\label{admis}
    \sigma(\tau) = \sigma_{n} = D_{0} + 2\sum^{\infty}_{k=2} \Rp \Big(D_{k}\frac{\tau^{k}}{R^k}\Big). 
\end{equation}

Using eqs.~\eqref{u_D_R}--\eqref{uepsR}, the expressions for displacements and stresses inside the disc are obtained, in terms of the traction coefficients given in eq.~\eqref{admis}, as
\begin{equation}
\label{finfieldsR}
    \begin{gathered}
    \frac{2\mu\, u(z)}{R} = 
    \sum^{\infty}_{k=2} D_k\frac{1}{k+1}\frac{z^{k+1}}{R^{k+1}} + 
    \sum^{\infty}_{k=2} \overline{D_k}\frac{{\overline{z}^{k-1}}}{R^{k-1}}\left(\frac{k}{k-1} - 
    \frac{z\overline{z}}{R^2}\right) + 2i\frac{z}{R} \sum^{\infty}_{k=2} \frac{1}{k^2-1} \Ip D_k, \\[3mm]
    \varepsilon_{11}(z)+\varepsilon_{22}(z)= 0, \\[3mm]
    \varepsilon_{22}(z)-\varepsilon_{11}(z)-2i\varepsilon_{12}(z) = 
    \frac{1}{\mu} \sum^{\infty}_{k=2} k\overline{D}_k\frac{{\overline{z}^{k-2}}}{R^{k-2}}\left(\frac{z\overline{z}}{R^2}-1\right), \\[3mm]
    p(z) = -\frac{\sigma_{11}(z)+\sigma_{22}(z)}{2} = 
    -\Rp{D_0}-2\Rp\sum^{\infty}_{k=2}D_k\frac{z^{k}}{R^k}, \\[3mm]
    \sigma_{22}(z)-\sigma_{11}(z)-2i\sigma_{12}(z) = 
    2\sum^{\infty}_{k=2}k\overline{D}_k\frac{{\overline{z}^{k-2}}}{R^{k-2}}\left(\frac{z\overline{z}}{R^2}-1\right).
\end{gathered}    
\end{equation}
%

It can be concluded from eqs.~\eqref{finfieldsR} that the deviatoric stress and all components of the strain tensor vanish at the disc boundary, where only the mean stress $p(z)$ differs from zero. The expressions for the remaining fields at the boundary are
\begin{equation}
    \begin{gathered}
    \frac{2\mu\, u(\tau)}{R} = \sum^{\infty}_{k=2} D_k\frac{1}{k+1}\frac{\tau^{k+1}}{R^{k+1}} + 
    \sum_{k=2}^{\infty} \overline{D}_k\frac{1}{k-1}\frac{R^{k-1}}{\tau^{k-1}} + 
    2i\frac{\tau}{R} \sum^{\infty}_{k=2} \frac{1}{k^2-1} \Ip D_k, \\[3mm]
    p(\tau) = -\frac{\sigma_{11}(\tau)+\sigma_{22}(\tau)}{2} = 
    -\Rp D_0 - 2 \Rp \sum_{k=2}^{\infty}D_k\frac{\tau^{k}}{R^k}. 
    \end{gathered}
\end{equation}

\subsection{The incompressible disc coated with an inextensible elastic rod}
\label{sec:Enforcement}

Models of coating for elastic solids, mimicking a thin and stiff linear elastic layer, have been idealized as elastic rods (in a 2D formulation) in \cite{benveniste1989stress, benveniste2001imperfect}. Later  \cite{zemlyanova2018circular,mogilevskaya2021fiber}, it was demonstrated that such coatings can be modeled as material surfaces possessing membrane stiffness \cite{gurtin1975continuum, GURTIN1978431}, enhanced with flexural and torsional stiffness, \cite{steigmann1997plane,steigmann1999elastic}. When the axial stiffness tends to infinity, these coating models impose an isoperimetric constraint on the disc. In the previous subsection, the displacement on the circle bounding the disc was assumed to satisfy the isoperimetric requirement, but without any additional effects. As proposed in \cite{gaibotti2023elastic}, the requirement can be enforced by coating the disc with an Euler rod, but in this way, unshearability and bending stiffness of the rod also play a role in the solution of the disc.  

The problem of an elastic disc coated with an inextensible elastic rod was analyzed in   \cite{gaibotti2023elastic} under the plane strain assumption settings without any restrictions on the value of Poisson's ratio. Moreover, effects of the prestress in the coating have been analyzed, leading to bifurcation \cite{gaibotti2024bifurcations} or to compliance effects \cite{gaibottiprestressato}. It can be concluded from the results obtained in Section \ref{mantegazza}, that all the developments presented in those articles remains valid for incompressible material deformed in plane strain, when $\nu=1/2$ is set. 

In \cite{gaibotti2023elastic}, the load on the Euler rod, coating the boundary of the disc, was expanded as 
(here we use different notations, namely coefficients $F_{\pm k}$ instead of coefficients $D_{\pm k}$ of \cite{gaibotti2023elastic})
\begin{equation}
\label{load}
    P\left(\tau\right) = \sum^{\infty}_{k=1} F_{-k}\frac{R^k}{\tau^{k}} + 
    \sum^{\infty}_{k=0} F_{k}\frac{\tau^k}{R^{k}}.
\end{equation}
The load $P$ is defined per unit length, so that in the notation of eq.~\eqref{sigma_sin} $P = b \sigma^{\star}$, where $b$ is the thickness of an out-of-plane portion of the disc. Similarly to eq.~\eqref{equilib}, equilibrium considerations lead to 
\begin{equation}
\label{eqload}
    \Ip F_{0} = F_{1} = 0.
\end{equation}

Imposing $\nu=1/2$ in eqs.~(45) and (47) of \cite{gaibotti2023elastic}, yields
\begin{equation}
    \begin{gathered}
    H_{2} = 0, \\[3mm]
    \frac{H_{1-k}}{R} = -\frac{(k+1)\bar{F}_{k} + 
    (k-1)F_{-k}}{4 k\mu b\,(k-1)\left[\frac{E_c}{24\mu}\left(\frac{h}{R}\right)^3 k(k^{2}-1)+1\right]}, 
    \quad k \geq 2, \\[3mm]
    \frac{H_{k+1}}{R} = -\frac{k+1)F_{k}+(k-1)\bar{F}_{-k}}{4k\mu b(k+1)\left[\frac{E_c}{24\mu}\left(\frac{h}{R}\right)^3 k(k^{2}-1)+1\right]}, 
    \quad k \geq 2,
    \end{gathered}
\end{equation}
where $E_c$ denotes the Young modulus of the beam (assumed with a rectangular cross-section of height $h$ and width $b$) coating the disc, which is characterized by a shear modulus $\mu$ and Young modulus $E_d$. 
Coefficients $D_k$ are given by 
\begin{equation}
    \begin{aligned}
    &D_{-1} = D_{1} = 0, \\[3mm]
    &\frac{D_{-k}}{2\mu} = \frac{-(k+1)\bar{F}_{k}-(k-1)F_{-k}}{4k\mu b\left[\frac{E_c}{24\mu}\left(\frac{h}{R}\right)^3k(k^{2}-1)+1\right]}, 
    \quad
    \frac{D_{k}}{2\mu} = \frac{-(k+1)F_{k}-(k-1)\bar{F}_{-k}}{4k\mu b\left[\frac{E_c}{24\mu}\left(\frac{h}{R}\right)^3k(k^{2}-1)+1\right]}, 
    \quad k \geq 2,
    \end{aligned}
\end{equation}
where in the notation of the present article $b=1$ and the load $P$ has to be understood as $\sigma^{\star}$, per unit area.

The above equations show that any load, expressible by eq.~\eqref{load} and satisfying eq.~\eqref{eqload}, except for the case in which $F_0$ and $F_{-1}$ are the only non-zero coefficients, provides admissible tractions on the boundary of the disc. However, the coefficient $D_0$ remains undetermined when the Poisson's ratio of the disc is $\nu=1/2$, while it vanishes for any other admissible value of $\nu$, see Table 2 in \cite{gaibotti2023elastic} where the latter coefficient is denoted as $B_0$. In fact, for any value of $\nu$, $D_0$ represents an internal uniform mean stress. 
This quantity remains undetermined when the disc is incompressible and deformed under plane strain. In this case, the uniform tractions generated on the disc’s boundary by its internal mean stress are absorbed by the coating and converted into an internal axial force. In other words, for an incompressible disc under plane strain coated by an inextensible rod, the stress state within the disc/coating system is determined up to an arbitrary, uniform, and isotropic residual stress.


\subsection{Nonlinear finite element simulations}

Non-trivial linear solutions exist even when both the incompressibility and isoperimetricity constraints are simultaneously enforced in a plane strain formulation of the disc/coating system. However, these solutions do not exist in a nonlinear formulation, as demonstrated numerically below.

Numerical simulations are carried out using Abaqus, in which the circular disc is discretized using $34800$ plane strain hybrid elements (CPE8RH) and $105041$ nodes, allowing for an accurate resolution of the deformation field in the nearly incompressible material. The disc is modelled with a Mooney-Rivlin hyperelastic constitutive law, characterized by material constants $C_{10}=\mu/2$, $C_{01}=0$, and $D_1=3(1-2\nu)/(\mu(1+\nu))$, ensuring that the initial stiffness matches that of a linear elastic material with shear modulus $\mu$ and Poisson’s ratio $\nu$. The coating, with thickness $h=R/10$ and applied along the perimeter of the disc, is modelled with hybrid beam elements (B23H), which incorporate both bending and axial behaviour. The mesh for the coating consists of $340$ elements distributed along the boundary. The TIE constraint is used to attach the coating to the disc. To simulate a nearly inextensible coating, the axial stiffness of the beam elements is set to $10^{10}\, E_d$, where $E_d$ is the elastic modulus of the disc, effectively constraining axial deformations. Mesh refinement is applied in the regions near the loaded surface to enhance accuracy in capturing the localized deformation and stress gradients. The applied load corresponds to the self-equilibrated distribution given by  eq.~\eqref{sigma_sin}. The peak load intensity $S_0$ in this distribution is set equal to $\mu_0 = E_d/2$, the initial shear modulus of a material with Young's modulus $E_d$ and null Poisson's ratio, $\nu=0$. To prevent rigid body motions, the node at the center of the disc is pinned, while a single point on the perimeter is constrained against circumferential displacement. The analysis is fully nonlinear, and geometric nonlinearities are accounted for by enabling the NLGEOM option in Abaqus, allowing for an accurate representation of large deformations and rotations. For comparison, linear analyses are also performed by turning off the NLGEOM option and using a static linear perturbation step, allowing the evaluation of the system’s response under the assumption of small deformations and linear material behaviour.

\begin{figure}[htb!]
    \centering
    \includegraphics[width=1.0\textwidth]{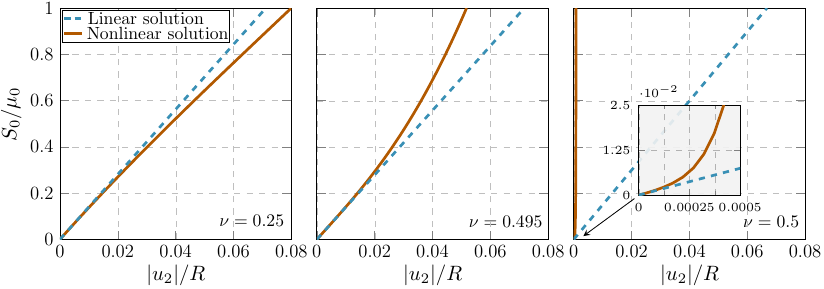}
    \caption{Normalized load $S_0/\mu_0$ as a function of normalized displacement at the center of the loaded region $|u_2|/R$, for three values of the Poisson’s ratio ($\nu=0.25$, $0.495$, and $0.5$). In the incompressible limit ($\nu=0.5$), the nonlinear solution displays locking behaviour, making the composite structure nearly undeformable, whereas the linear solution remains compliant. The structure is not fully undeformable because the beam coating the disc retains a small but finite axial compliance.}
    \label{fig:locking}
\end{figure}

The load-displacement curves, evaluated at $\alpha=\pi/2$, are presented in Fig.~\ref{fig:locking}, for three values of the Poisson’s ratio: $0.25$, $0.495$, and $0.5$. The dimensionless maximum load, $S_0/\mu_0$, is reported as a function of the dimensionless displacement $|u_2|/R$, where $u_2$ is the vertical displacement at the center of the loaded region, and $R$ is the disc radius. 

The results show that at $\nu = 0.25$, the nonlinear solution is more compliant than the linear solution. However, this trend reverses as $\nu$ increases, and in the incompressible limit ($\nu = 0.5$), the nonlinear solution exhibits locking behavior. Consequently, the composite structure becomes effectively undeformable (the results still show some deformability because the coating beam is not fully axially inextensible). In contrast, the linear solution predicts a compliant response even in the incompressible limit, revealing a fundamental inconsistency of the linear analysis in capturing the behavior of the structure under extreme material constraints.

\begin{figure}[htb!]
    \centering
    \includegraphics[scale=1.1,keepaspectratio]{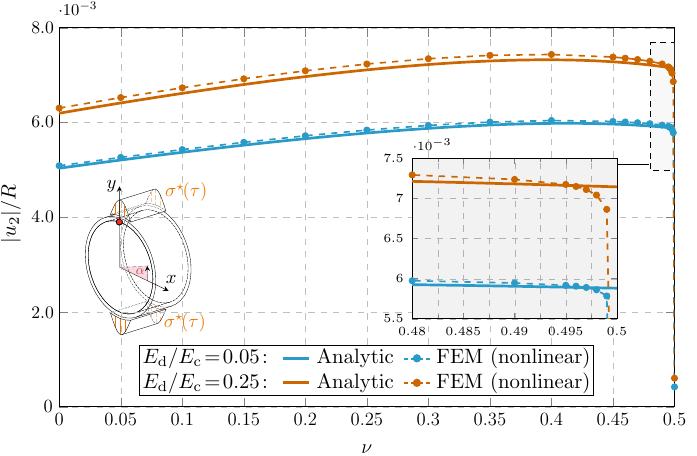}
    \caption{Dimensionless boundary displacement $|u_2|/R$ evaluated at $\alpha = \pi/2$ for an elastic disc coated with an inextensible elastic rod (height $h = R/10$), as a function of the Poisson's ratio $\nu$ of the disc material. The extrados of the rod is subjected to a sinusoidal traction $\sigma^{\star}(\tau)$, centered at $\alpha_0 = \pm \pi/2$ and symmetrically distributed over an angle $\gamma = 10^{\circ}$. The load magnitude is set to $S_0 = E_d/20$, and two different Young's modulus ratios $E_d/E_c$ (disc to coating) are considered. Nonlinear numerical results are shown as discrete points, together with the linear analytical solution (continuous lines). The nonlinear numerical results show that the displacement tends to vanish as incompressibility is approached ($\nu \to 1/2$), whereas the linear solution predicts a finite displacement in this limit.}
    \label{u2_tau_plot}
\end{figure}

\begin{figure}[htb!]
    \centering
    \includegraphics[scale=1.1,keepaspectratio]{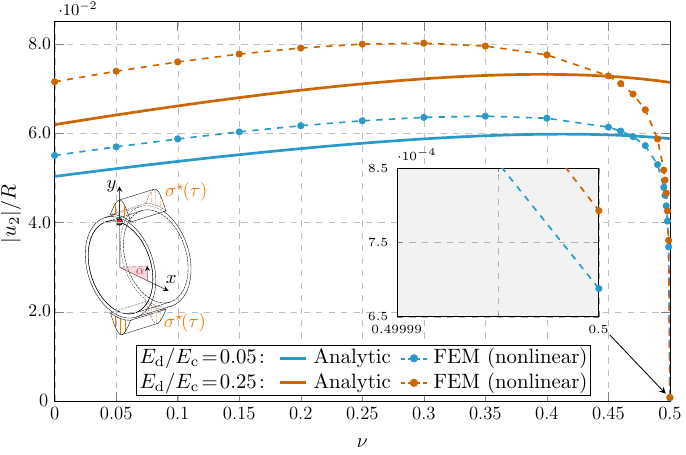}
    \caption{As in Fig. 4, except that the applied load magnitude is significantly increased to $S_0 = E_d/2$ to highlight the effects of large strains. The nonlinear analysis shows higher compliance than the linear prediction at small values of $\nu$, while, as $\nu \to 1/2$, the composite structure progressively locks and behaves as a rigid body, causing the displacement to vanish.}
    \label{u2_tau_plot2}
\end{figure}

Figures~\ref{u2_tau_plot} and \ref{u2_tau_plot2} present the normalized displacement $|u_2|/R$, evaluated at $\alpha = \pi/2$ and corresponding to the final applied load, plotted as a function of the Poisson’s ratio $\nu$ of the disc material. Two values of the ratio between the Young’s moduli of the disc and the coating are considered, namely $E_d/E_c = \{0.05, 0.25\}$. For a relatively small applied load, $S_0 = E_d/20$ (see Fig.~\ref{u2_tau_plot}), the numerical results from the nonlinear analysis (dots) and the analytical linear solutions (solid lines) show excellent agreement, with discrepancies confined to a narrow region near $\nu = 0.5$, where the nonlinear solutions collapse to $u_2 = 0$. A higher load, $S_0 = E_d/2$, is considered in Fig.~\ref{u2_tau_plot2} to highlight the differences between the linear and nonlinear responses.

Figures~\ref{u2_tau_plot} and, even more clearly, Fig.~\ref{u2_tau_plot2} show that as the Poisson's ratio $\nu$ increases, the nonlinear solution transitions from being more compliant to being stiffer than the linear one. In the incompressible limit $\nu \to 0.5$, the linear solution approaches a non-trivial deformation state, whereas the nonlinear solution tends toward the undeformed configuration.

\begin{figure}[htb!]
    \centering
    \includegraphics[width=1.0\linewidth]{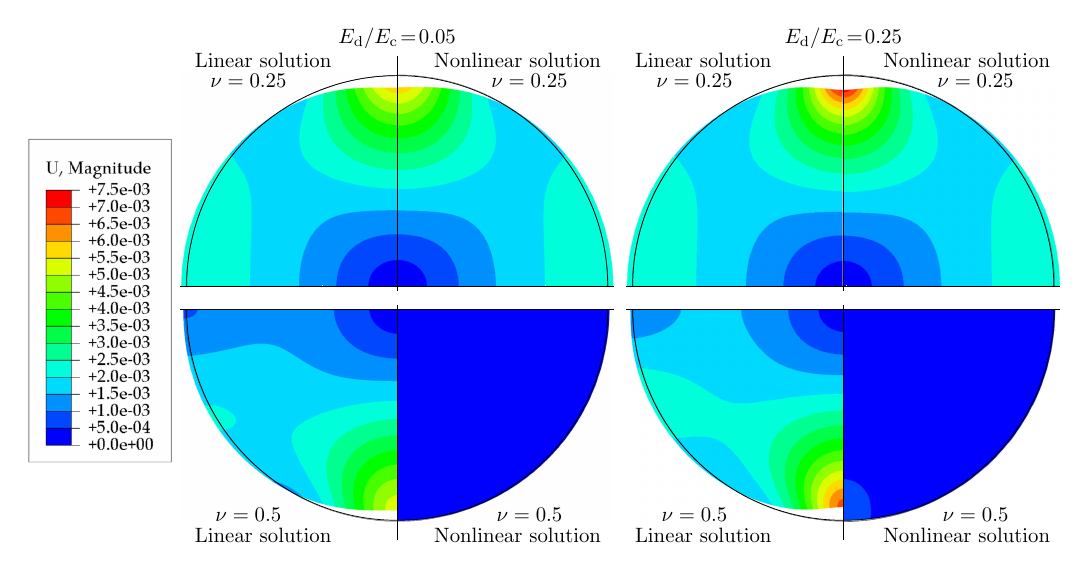}
    \caption{Maps of displacement magnitude (magnified by a factor of $10$) for the deformed configurations of the disc/coating structure under a load of $S_0 = E_d/20$ and for $E_d/E_c = \{0.05, 0.25\}$. Both linear and nonlinear solutions, obtained via FEM analysis, are shown side by side for comparison. Two values of Poisson's ratio are considered: $\nu = 0.25$ (upper panels) and $\nu = 0.50$ (lower panels). The linear solution closely matches the nonlinear one for $\nu = 0.25$. However, in the incompressible limit $\nu = 0.5$, the linear solution significantly deviates and shows a compliant response, while the nonlinear solution correctly predicts nearly undeformable behavior (the small deformation is due to the axial stiffness of the coating beam, which is high but not infinite).}
    \label{fig:disp}
\end{figure}

The same conclusion drawn from the previous figures can also be inferred from Fig.~\ref{fig:disp}, which presents maps of the displacement magnitude plotted on the deformed coated disc system.

The solution exhibits two axes of symmetry, so that only a quarter of the domain is shown. The linear solution (on the left) and the nonlinear solution (on the right) are displayed side by side for easier comparison. Two ratios of Young’s moduli, $E_d/E_c=\{0.05,0.25\}$, are considered, along with two values of Poisson’s ratio: $\nu = 0.25$ and the incompressible limit $\nu = 0.5$. The deformation of the disc is computed with $S_0 = E_d/20$, and the result is magnified by a factor of $10$ for better visualization. It is evident that the nonlinear and linear analyses are nearly identical at $\nu = 0.25$. However, the nonlinear solution exhibits a rigid response when $\nu = 0.5$.

\subsection{Bifurcation of the disc coated with a rod}

Subjected to a uniformly distributed pressure $\Pi(n)$, an elastic disc coated with an axially inextensible rod may bifurcate for a sufficiently high load value. Note that the disc remains unloaded and the rod is subjected to a purely axial load before bifurcation. This problem was investigated for a compressible disc in \cite{gaibotti2024bifurcations}. However, that solution remains valid for incompressible materials as $\nu \to 1/2$, when a state of plane strain prevails. A full generality was assumed in \cite{gaibotti2024bifurcations}, where bifurcation loads were determined for three different types of uniform radial tractions applied to the coating, and complete bonding or slip bonding at the coating/disc interface. 

When $\nu=1/2$, the bifurcation radial load $\Pi(n)$ corresponding to the $n$-th bifurcation mode is still expressed by Eq.~(4.39), derived and reported in \cite{gaibotti2024bifurcations}, as a particular case. In the present notation, the bifurcation radial loads can be expressed as 
\begin{equation}
\label{pcr_nu05}
    \frac{\Pi(n)R^3}{E_c b h^3} = 
    \frac{n^2(n^2-1) + 8n\frac{E_d}{E_c}
    \big(\frac{R}{h}\big)^3}{12(n^2-1)+6\xi\left[(1-n)^{\beta-1}+(1+n)^{\beta-1}\right]},
    \quad n \geq 2,
\end{equation}
where the coefficients $\xi$ and $\beta$ describe the nature of the applied radial load. In particular $\xi=\beta=1$ for hydrostatic pressure, $\xi=1$, $\beta=0$ for centrally directed load and $\xi=\beta=0$ for dead load. 

\begin{figure}[htb!]
    \centering
    \includegraphics[scale=1.1,keepaspectratio]{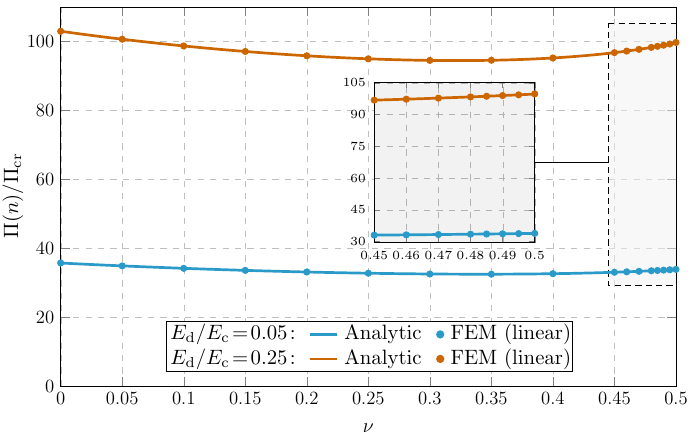}
    \caption{Normalized critical pressure $\Pi(n)/\Pi_{cr}$ for bifurcation as a function of the Poisson's ratio of the disc,  when the latter is coated with an elastic rod subject to a hydrostatic pressure distribution over the boundary, $\xi=\beta=1$ in Eq. \eqref{pcr_nu05}. Analytical results (solid lines), Eq.~(\ref{pcr_nu05}) are compared with the FEM simulation (dots) for $E_{d}/E_{c}=\left\{0.05,0.25\right\}$. Only the first mode, $n=1$, is investigated.  
    Results are reported for a rod with a rectangular cross-section with a height $h=R/10$. Note that the buckling pressure converges for $\nu \to 1/2$ to a finite value, while the fully nonlinear analyses presented in Figs.~\ref{u2_tau_plot} and \ref{u2_tau_plot2} show that bifurcation cannot develop in this limit.}
    \label{pcr_nu_plot_hydro}
\end{figure}

Details regarding the nature of these types of pressure loads can be found in \cite{boresi1955refinement,bodner1958conservativeness, gaibottianello, gaibotti2024bifurcations, gaibottiprestressato}. In Fig.~\ref{pcr_nu_plot_hydro}, the critical pressure $\Pi(n)$ is reported as a function of the disc Poisson's ratio $\nu$. The coated disc is assumed to be subjected to a hydrostatic pressure distribution ($\xi=\beta=1$). Results are presented for two values of the ratio $E_d/E_c = \{0.05,0.25\}$ and made dimensionless through division by the critical radial load for the buckling of the annular beam,  considered isolated from the disc ($\Pi_{\text{cr}} = E_c b h^3/4R^3$). 
Note that in the limit of vanishing coating stiffness, $E_ch^3/E_dR^3 \to 0$, the critical load tends to zero for a bifurcation mode involving an infinite number of creases, $n \to \infty$.

The analytical results, shown as solid lines in Fig.~\ref{pcr_nu_plot_hydro}, exhibit excellent agreement with the {\it linearized} FEM solutions, represented by dots in the same figure. The numerical simulations are performed using Abaqus in a setting similar to that previously described. The circular disc is discretized with  $28018$ biquadratic plane strain elements with a hybrid formulation (CPE8RH), while the coating is modeled with $200$ quadratic beam elements, also employing a hybrid formulation (B22H). The buckling analysis is performed using the linear perturbation buckle step, which solves the corresponding generalized eigenvalue problem. 

The key observation from Fig.~\ref{pcr_nu_plot_hydro} is that the linear numerical solution aligns precisely with the analytical prediction, yielding a finite bifurcation load in the incompressible limit $\nu \to 1/2$, as given by eq.~\eqref{pcr_nu05}. It is worth noting that, when a finite element analysis is performed in the incompressible limit $\nu = 1/2$, the standard buckling procedure, based on linearization, still returns the same finite bifurcation load shown in Fig.~\ref{pcr_nu_plot_hydro}. Hence, the bifurcation analysis of a coated disc gives an example of a discontinuous limit: as $\nu \to 1/2$, the actual buckling load jumps to infinity, a behavior that linearized governing equations cannot capture.

\section{The \lq bathyscaphe lesson', or the importance of the linear solution}

As explained at the beginning of the previous section, the fact that non-trivial linear solutions exist for an (i.) incompressible disc, (ii.) deformed under plane strain, and (iii.) coated with an isoperimetric constraint is a consequence of the linearization of the mechanical response of the system under analysis. 

\begin{figure}[htb!]
    \centering
    \includegraphics[scale=0.25,keepaspectratio]{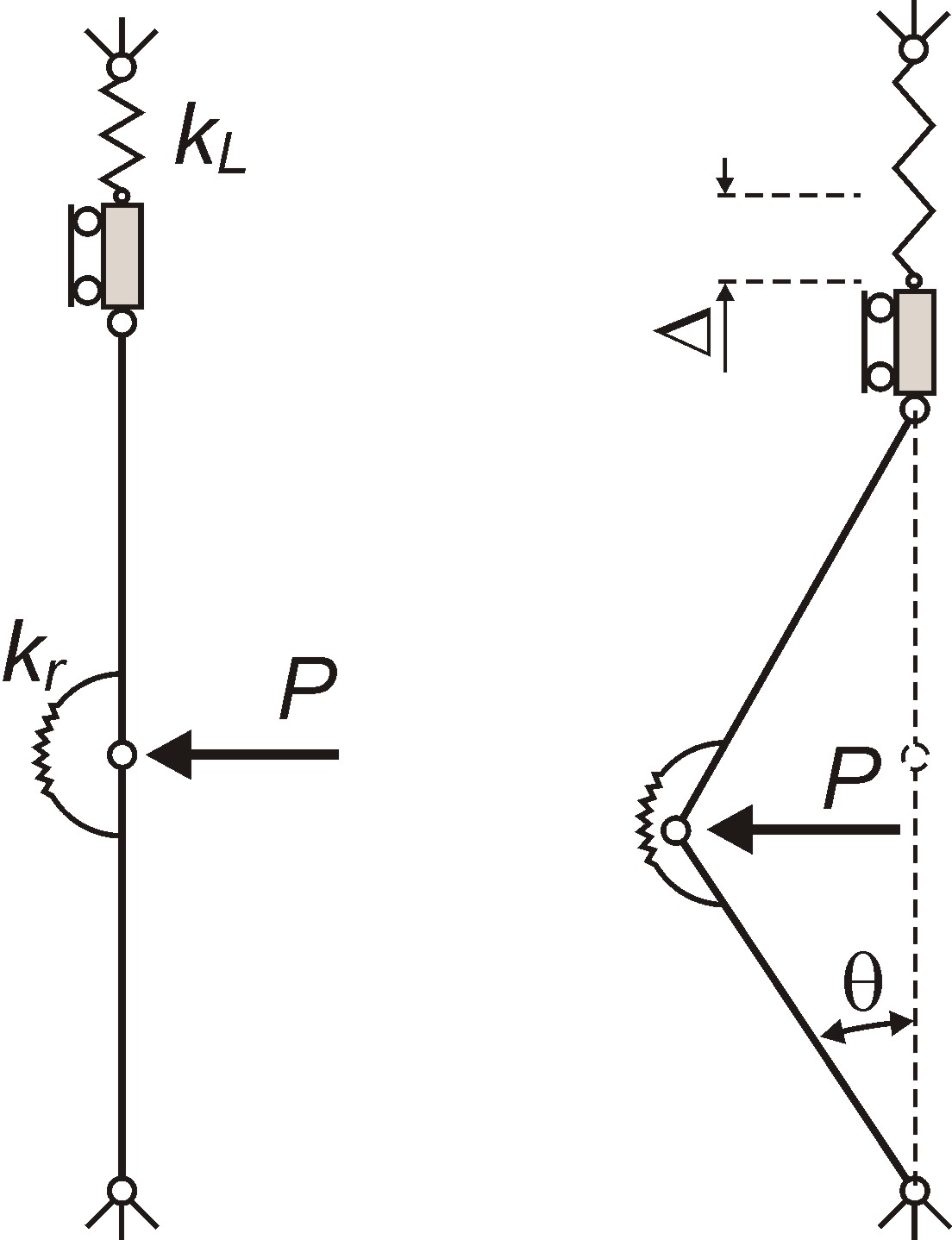}
    \caption{A single d.o.f structure in its undeformed (left) and deformed (right) configurations, loaded with a transverse force $P$. All the bars are rigid, one hinge is elastic, with stiffness $k_r$, and a longitudinal spring of stiffness $k_L$ is applied on a roller. The axial elongation $\Delta$ of the spring $k_L$ is essential for deformation.}
    \label{buck}
\end{figure}

The situation is strictly similar to that illustrated in Fig.~\ref{buck}, where a structure with a single degree of freedom $\theta$ is composed of two rigid rods (of length $l$) joined through an elastic hinge (of stiffness $k_r$). The two rods are externally constrained with a fixed pivot (at the lower end) and with a roller subject to the action of an extensional spring (of stiffness $k_L$). 

The structure is loaded with a force $P$ applied transversely on the elastic hinge. The total potential elastic energy is 
\begin{equation}
    \mathcal{P} = 2 k_r \theta^2 - Pl \sin\theta + 2 k_L l^2 \big(1-\cos\theta\big)^2,
\end{equation}
so that equilibrium is obtained by deriving with respect to $\theta$ 
\begin{equation}
\label{soc}
    \frac{d\,\mathcal{P}}{d\theta} = 4k_r \theta - Pl \cos\theta + 
    4 k_L l^2 \sin\theta \big(1-\cos\theta\big) = 0, 
\end{equation}
an equation which can be expanded in a Taylor series about $\theta=0$ as
\begin{equation}
\label{mel}
    \frac{d\,\mathcal{P}}{d\theta} = -Pl + 4k_r \theta + 
    \frac{Pl}{2} \theta^2 + 2k_L l^2 \theta^3 + O(\theta^4) = 0. 
\end{equation}

Equation~\eqref{mel} shows that, as $\theta$ tends to zero, the well-known linear solution is obtained, $\theta=Pl/(4k_r)$, which is independent of the stiffness $k_L$, entering the equilibrium equation only at the third-order term. On the other hand, eq.~\eqref{soc} shows that as $K_L \to \infty$, only the solution $\theta=0$ becomes possible. This is analogous to the incompressible disc with isoperimetric constraint, where a linear solution remains possible, even when a nonlinear solution would only provide the trivial result. 

\begin{figure}[htb!]
    \centering
    \includegraphics[scale=0.25,keepaspectratio]{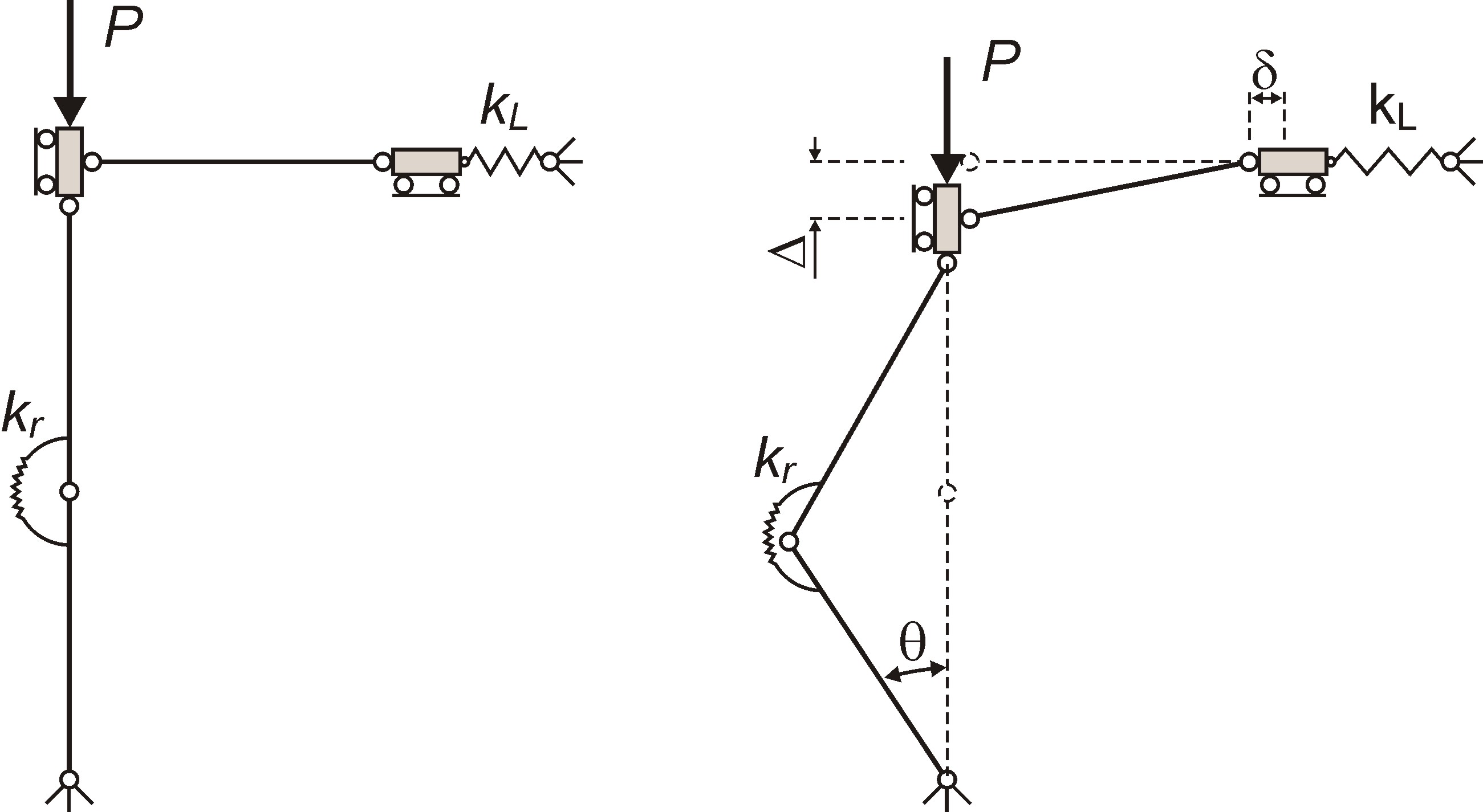}
    \caption{A single d.o.f structure, loaded with an axial force $P$ in its undeformed (left) and deformed (right) configurations. All the bars are rigid, one hinge is elastic, with stiffness $k_r$, and a longitudinal spring of stiffness $k_L$ is applied on a slider. The axial elongation $\Delta$ of the spring $k_L$ is essential for buckling.}
    \label{buck_2}
\end{figure}

In the buckling problem for the coated disc analyzed in \cite{gaibotti2024bifurcations} and reported in the previous section, a finite value of buckling pressure is obtained in the limit $\nu \to 1/2$, although the plane strain assumption and the isoperimetric constraint are both enforced, which would imply an infinite buckling load. This behavior is again related to the linearization applied to the bifurcation problem 
and can be explained with the structure shown in Fig.~\ref{buck_2}. 

The structure closely resembles the one shown in Fig.~\ref{buck}, except that the load is vertical and applied on the roller, which is constrained with a horizontal rigid rod, terminating with another roller equipped with the longitudinal spring of stiffness $K_L$. 

The load $P$ compresses only the two vertical rods in the trivial solution, while in a non-trivial deformation, the total potential energy can be found by defining the two displacements 
\begin{equation}
    \Delta = 2l - 2l \cos\theta, 
    \quad
    \delta = l - l \cos \arcsin \big(2 - 2 \cos\theta \big), 
\end{equation}
in the form 
\begin{equation}
\label{piaz}
    \mathcal{P} = 2 k_r \theta^2 - 2 Pl (1 - \cos\theta) + 
    \frac{k_L l^2}{2} \bigg[1 - \cos \arcsin \big(2 - 2 \cos\theta \big)\bigg]^2. 
\end{equation}

A derivative of eq.~(\ref{piaz}) with respect to $\theta$ and a Taylor series expansion of the result truncated at 7-th order leads to the equilibrium equation as 
\begin{equation}
\label{ban}
    P = \frac{2520\big(4k_r + k_Ll^2 \theta^6\big)}{l\big(5040 - 840\theta^2 + 42\theta^4 - \theta^6\big)}. 
\end{equation}

Equation~\eqref{ban} shows that the linearized solution provides the finite buckling load $P_{cr}=2k_r/l$, which loses validity when $k_L \to \infty$, correctly providing an infinite value for the buckling load. 

At this stage, it might seem to the reader that the linear solution is meaningless, so that only the nonlinear solution correctly predicts the trivial result. This is not true, as the linearized solution still retains importance, from the theoretical point of view and, perhaps more importantly, from an engineering perspective. 

Theoretically, the linearized solution may provide a stress distribution in a system that behaves as a rigid body. Although this distribution remains undetermined in that case, the linearized solution offers an approximation that aligns with attempts to define it nonetheless \cite{grioli, royer_10, royer_11}. 

The interest in the linearized solution for design can readily be appreciated from the bifurcation solution, in which a finite value for the buckling load is smoothly approached when $\nu$ reaches the value $1/2$, Fig.~\ref{pcr_nu_plot_hydro}. This buckling load remains valid until the limit value is attained, in the same sense that the value $P_{cr}=2k_r/l$ remains valid for any large but finite value of $k_L$. 
In reality, perfect constraints do not exist, and any real structure is expected to exhibit some problem when a buckling load obtained with a linearization is attained. For a bifurcated mode, it is certainly harder to develop in the presence than in the absence of even imperfect constraints, but the initiation of buckling will always be marked by some structural \lq symptom', which may vary in severity.
We propose to nickname this effect the \lq bathyscaphe lesson', taking inspiration from the account of the Trieste bathyscaphe descent into the Mariana Trench, reporting on a loud noise heard by the crew at approximately 9450~m depth. In fact, the bathyscaphe was designed following the same principle as a free balloon moving in air. Thus, the float of the bathyscaphe was filled with a buoyant substance, which was selected to be aviation gasoline, because of its low compressibility. The latter feature was essential to maintain buoyancy under the increasing pressure of the external water, but also to provide a nearly incompressible core inside a thin steel float vessel, to prevent buckling under external pressure. Therefore, the design of the bathyscaphe resembles the coated incompressible disc, so that the loud noise might be interpreted as caused by a buckling, which could not develop further due to the constraints. It should be noted that the presented analogy is intended to provide only a nickname to the effects that should be expected when a bifurcation load, impossible for nonlinear behavior, but possible in the linearized solution, is encountered, because the float of the Trieste was carefully pressure-compensated with a two-way \lq breathing' valve \cite{batiscafo}.

\section{Concluding remarks}
\label{sec:concluding}

Two-dimensional linear elastic solutions under the isochoric constraint have been explored. This is an interesting case for several reasons, and also due to the formal equivalence of the governing equations with those of slow viscous flow in a Newtonian fluid. Using complex potentials, an analytical solution is derived for a circular disc subjected to arbitrary, self-equilibrated traction distributions along its boundary. The analysis is then extended to include an isoperimetric constraint on the boundary. Although the body, under such double constraints, is expected to behave as a rigid solid and remain undeformed under any load distribution, it is demonstrated that non-trivial linearized solutions still exist. These solutions, however, are ruled out when a fully nonlinear analysis is carried out, as shown through a numerical example.

While the linear solutions might be regarded as a limitation of the linearized theory, a less superficial interpretation is that they retain relevance -- both theoretically, as representations of stress states within a rigid body, and practically, in engineering contexts where ideal constraints are never perfectly realized.

\section*{Acknowledgements}
D.B. gratefully acknowledges financial support from ERC-ADG-2021-101052956-BEYOND. A.P. gratefully acknowledges funding from the European Union (ERC-CoG-2022-101086644-SFOAM). S.G.M. gratefully acknowledges support from National Science Foundation, USA, award NSF CMMI-2112894. M.G. gratefully acknowledges funding from the European Union - Next Generation EU, in the framework of the PRIN 2022 project n. 2022NNTZNM “Unveiling embodied intelligence in natural systems for bioinspired actuator design” (CUP: G53D23001240006). D.B., A.P., and M.G. are members of the ‘Gruppo Nazionale di Fisica Matematica’ (GNFM) of the ‘Istituto Nazionale di Alta Matematica’ (INdAM).



\appendix

\section*{Appendix}
\label{sec:appendix}
\input{axial_extensibility.tex}

{\newpage
\renewcommand{\bibname}{References}
\printbibliography
}

\end{document}

%% file: definitions.tex
\DeclareMathOperator{\tr}{tr}

\DeclareMathOperator{\diver}{div}

\DeclareMathOperator{\Rp}{Re}
\DeclareMathOperator{\Ip}{Im}

\newcommand{\bzero}{\bm 0}

\newcommand{\be}{\bm e}

\newcommand{\bem}{\bm m}
\newcommand{\bn}{\bm n}

\newcommand{\bs}{\bm s}

\newcommand{\bu}{\bm u}

\newcommand{\bx}{\bm x}

\newcommand{\bI}{\bm I}

\newcommand{\bvarepsilon}{\bm \varepsilon}

\newcommand{\bsigma}{\bm \sigma}



%% file: axial_extensibility.tex
\section{Axial inextensibility of the circular rod}
\label{App_extensibility}

A circular two-dimensional rod (belonging to the plane $(x,y)$, spanned by the base vector $\be_1$, $\be_2$) is parameterized through the arc lengths $s_\tau$ and $s_t$, defined in the reference and current configurations respectively. The tangents ${\bs}_0$ and ${\bs}$ at a point in the two configurations are 
\begin{equation}
\label{dtdt}
    {\bs}_0 = \frac{\partial {\bx}_0}{\partial s_\tau}, 
    \quad 
    {\bs} = \frac{\partial{\bx}}{\partial s_t},
\end{equation}
where $\bx_0$ and $\bx$ are the position vectors of a point in the reference and current configuration, respectively. Moreover, the vector $\bm{m}_0 = \bm{s}_0 \times \bm{e}_3$, ($\bm{e}_3 = \bm{e}_1 \times \bm{e}_2$), is defined, radial and pointing outward from the center of the reference configuration.
The displacement $\bu$ brings a point at $\bx_0$ in the reference configuration to the point $\bx$ in the current configuration, so that the displacement and the position vector $\bx$ introduced in eq.~\eqref{dtdt}$_2$ can be written as 
\begin{equation}
\label{urr}
    \bu = u_{\alpha} \bs_0 + u_{r} \bem_0, 
    \quad 
    \bx = \bx_0 + \bu.
\end{equation}
On application of the chain rule and noting that the ratio $\partial s_t/\partial s_\tau = \lambda$ represents the axial stretch, eq.~\eqref{dtdt}$_2$ yields
\begin{equation}
\label{dsds0}
    \frac{\partial{\bx}}{\partial s_t} = 
    \frac{\partial{\bx}_0}{\partial s_t} + \frac{\partial{\bu}}{\partial s_t} = 
    \frac{1}{\lambda}\Big({\bs}_0 + \frac{\partial{\bu}}{\partial s_\tau}\Big),
\end{equation}
where, from eq.~\eqref{u} it follows
\begin{equation}
\label{du}
    \frac{\partial{\bu}}{\partial s_\tau} = 
    \frac{\partial u_\alpha}{\partial s_\tau}{\bs}_0 + u_\alpha\frac{\partial {\bs}_0}{\partial s_\tau} + 
    \frac{\partial u_r}{\partial s_\tau}{\bem}_0 + u_r\frac{\partial {\bem}_0}{\partial s_\tau}.
\end{equation}
Eqs.~(3.2), derived and reported in \cite{gaibotti2024bifurcations}, is expressed in the present notation as 
\begin{equation}
\label{geomdt}
    \frac{\partial{\bs}_0}{\partial s_\tau} = -\frac{1}{R}{\bem}_0, 
    \quad 
    \frac{\partial {\bem}_0}{\partial s_\tau}=\frac{{\bs}_0}{R},
\end{equation}
so that it follows from eq.~\eqref{du}
\begin{equation}
\label{du2}
    \frac{\partial{\bu}}{\partial s_\tau} = 
    \Big(\frac{\partial u_{\alpha}}{\partial s_\tau} + 
    \frac{u_r}{R}\Big){\bs}_0+\Big(\frac{\partial u_r}{\partial s_\tau} - 
    \frac{u_{\alpha}}{R}\Big){\bem}_0.
\end{equation}

The inextensibility condition of the rod ($\lambda=1$) implies that $||\partial {\bx}/\partial s_t||^2=1$ in eq.~\eqref{dsds0}, which leads to 
\begin{equation}
\label{norm_S}
   2\frac{\partial{\bu}}{\partial s_\tau}\cdot{\bs}_0 + 
   \frac{\partial {\bu}}{\partial s_\tau}\cdot\frac{\partial {\bu}}{\partial s_\tau} = 0. 
\end{equation}

A substitution of eq.~\eqref{du2} in eq.~\eqref{norm_S}, plus the relation $\partial s_\tau/\partial\alpha=R$, 
leads to eq.~\eqref{polpettone}.